\documentclass[submission,copyright,creativecommons]{eptcs}
\providecommand{\event}{MSFP 2022} 
\usepackage{breakurl}             
\usepackage{underscore}           

\title{Tableless Calculation of Circular Functions on Dyadic Rationals}
\author{Peter Kourzanov
  \institute{TU Delft}
  \email{kourzanov@acm.org}
}
\def\titlerunning{Tableless Trigonometry on Dyadic Rationals}
\def\authorrunning{P. Kourzanov}

\usepackage{hyperref}
\usepackage{color}
\usepackage{graphicx}
\usepackage[utf8]{inputenc}
\usepackage{acronym}
\usepackage{multicol}
\usepackage{amsmath} 
\usepackage{amsfonts}
\usepackage{floatflt}
\usepackage{tikz}
\usetikzlibrary{trees}
\def\nofootnote#1{}

\def\cmd#1{{\tt #1}}

\def\describe#1#2{\item[#1]\hfill\parbox[t]{.75\columnwidth}{\raggedleft #2}
	\index{#1!explanation of}}

\def\macro#1#2{\acrodef{#1}{#2}\describe{#1}{#2}}
\def\macrod#1#2#3{\acrodef{#1}{#2}\describe{#1}{#2 (#3)}}
\def\mac#1{\ac{#1}\index{#1}}

\def\macs#1{\acs{#1}\index{#1}}

\def\cref#1{\ref{ch:#1}}

\newsavebox{\fm}

{%
	\end{minipage}%
	\end{lrbox}%
	\fbox{\usebox{\fm}}%
}

\newsavebox{\capt}

\newsavebox{\alldefs}

\newcommand{\HFigure}[3][1]{%
\begin{figure}[htb]%
\def\lab{\label{fig:#3}}%
\ifx#2\empty\else\caption{#2}\fi%
\centering\resizebox*{#1\textwidth}{!}{\includegraphics{#3}}%
\lab%
\end{figure}%
}

\newcommand{\hFigureSS}[3][1]{%
\vspace{-1em}
\begin{figure}[htb]%
\def\lab{\label{fig:#3}}%
\centering\resizebox*{#1\columnwidth}{!}{\includegraphics{#3}}%
\ifx#2\empty\else%
\vspace{-1em}
\caption{#2}\fi%
\vspace{-1em}
\lab%
\end{figure}%
}

\newcommand{\hFigureSR}[3][1]{%
\vspace{-1em}
\begin{figure}[htb]%
\def\lab{\label{fig:#3}}%
\hspace{-3em}\resizebox*{#1\columnwidth}{!}{\rotatebox{-90}{\includegraphics{#3}}}
\ifx#2\empty\else%
\caption{#2}\fi%
\lab%
\end{figure}%
}

\newcommand{\hFigureS}[3][1]{%
\vspace{-1em}
\begin{figure}[htb]%
\def\lab{\label{fig:#3}}%
\centering\resizebox*{#1\columnwidth}{!}{\includegraphics{#3}}%
\ifx#2\empty\else%
\vspace{-1em}
\caption{#2}\fi%
\lab%
\end{figure}%
}

\newcommand{\hFigure}[3][1]{%
\begin{figure}[htb]%
\def\lab{\label{fig:#3}}%
\centering\resizebox*{#1\textwidth}{!}{\includegraphics{#3}}%
\ifx#2\empty\else\caption{#2}\fi%
\lab%
\end{figure}%
}

\newcommand{\VFigure}[3][1]{%
\begin{figure}[htb]%
\def\lab{\label{fig:#3}}%
\ifx#2\empty\else\caption{#2}\fi%
\centering\resizebox*{!}{#1\textheight}{\includegraphics{#3}}%
\lab%
\end{figure}%
}

\newcommand{\vFigure}[3][1]{%
\begin{figure}[htb]%
\def\lab{\label{fig:#3}}%
\centering\resizebox*{!}{#1\textheight}{\includegraphics{#3}}%
\ifx#2\empty\else\caption{#2}\fi%
\lab%
\end{figure}%
}

\newcommand{\vFigureR}[3][1]{%
\begin{figure}[htb]%
\def\lab{\label{fig:#3}}%
\centering\resizebox*{!}{#1\textheight}{\rotatebox{270}{\includegraphics{#3}}}%
\ifx#2\empty\else\caption{#2}\fi%
\lab%
\end{figure}%
}

\newcommand{\HFFigure}[3][1]{%
\begin{floatingfigure}[v]{#1\textwidth}%
\def\lab{\label{fig:#3}}%
\ifx#2\empty\else\caption{#2}\fi%
\resizebox*{#1\textwidth}{!}{\includegraphics{#3}}%
\lab%
\end{floatingfigure}%
}

\newcommand{\hFFigure}[3][1]{%
\begin{floatingfigure}[v]{#1\textwidth}%
\def\lab{\label{fig:#3}}%
\resizebox*{#1\textwidth}{!}{\includegraphics{#3}}%
\ifx#2\empty\else\caption{#2}\fi%
\lab%
\end{floatingfigure}%
}

\newcommand{\VFFigure}[3][1]{%
\begin{floatingfigure}[v]{#1\textwidth}%
\def\lab{\label{fig:#3}}%
\ifx#2\empty\else\caption{#2}\fi%
\resizebox*{!}{#1\textheight}{\includegraphics{#3}}%
\lab%
\end{floatingfigure}%
}

\newcommand{\vFFigure}[3][1]{%
\begin{floatingfigure}[v]{#1\textwidth}%
\def\lab{\label{fig:#3}}%
\resizebox*{!}{#1\textheight}{\includegraphics{#3}}%
\ifx#2\empty\else\caption{#2}\fi%
\lab%
\end{floatingfigure}%
}

\newcommand{\HLFigure}[4][1]{%
\begin{figure}[htb]%
\def\lab{\label{fig:#3}}%
\ifx#2\empty\else\caption{#2}\fi%
{\hfill%
#4%
\hfill}%
\lab%
\end{figure}%
}

\newcommand{\hLFigure}[4][1]{%
\begin{figure}[htb]%
\def\lab{\label{fig:#3}}%
{\hfill%
#4%
\hfill}%
\ifx#2\empty\else\caption{#2}\fi%
\lab%
\end{figure}%
}

\def\CEE/{\textsc{C}\spacefactor 1000 }
\def\CPP/{\textsc{C++}\spacefactor 1000 }
\def\CWEB/{{\tt CWEB\/}}
\def\SPARC{SPARC\-\kern.1em station}
\def\UNIX/{{\small UNIX\/}}
\def\WEB/{{\tt WEB\/}}

\def\PresentationOnly#1{%
  \ifx\everyslide\empty%
     #1%
  \fi%
}

\def\DocumentationOnly#1{%
  \ifx\everyslide\empty\else%
     #1%
  \fi%
}
\def\EitherDoc#1#2{%
  \ifx\everyslide\empty%
     #1%
  \else%
     #2%
  \fi%
}

\def\tlead{%
  \xleaders\hbox to 10pt{\hfill.\hfill}%
}

\def\dotfill#1{%
  \hbox to #1{\tlead\hfill}%
  \tlead\hfil%
}

\def\otfill{%
  \hbox to 20pt{\tlead\hfill}%
  \tlead\hfil%
}

\def\vdownarrowfill{%
  \hbox{$\mkern20mu|$}%
  \kern-2pt%
  \cleaders\hbox{$\mkern20mu|$}\vfill%
  \kern-4pt%
  \hbox{$\mkern18.1mu\downarrow$}%
}

\def\vuparrowfill{%
  \hbox{$\mkern18.1mu\uparrow$}%
  \kern-2pt%
  \cleaders\hbox{$\mkern20mu|$}\vfill%
  \kern-4pt%
  \hbox{$\mkern20mu|$}%
}

\def\Fboxhskip{3pt}
\def\Fboxvskip{3pt}

\def\Fbox#1{%
  \hbox{\lower\Fboxvskip\vbox{\hbox{%
  \vrule%
  \vbox{%
    \hrule%
    \kern\Fboxvskip%
    \hbox{%
      \kern\Fboxhskip%
      #1%
      \kern\Fboxhskip%
    }%
    \kern\Fboxvskip%
    \hrule%
  }%
  \vrule%
  }}}%
}

\def\Ebox#1{%
  \hbox{\lower\Fboxvskip\vbox{\hbox{%
  \vbox{%
    \kern\Fboxvskip%
    \hbox{%
      \kern\Fboxhskip%
      #1%
      \kern\Fboxhskip%
    }%
    \kern\Fboxvskip%
  }%
  }}}%
}

\def\rev#1#2{%
  \ifx#2.%
    #1%
    \def\next{\relax}%
  \else%
    \def\next{\rev{#2#1}}%
  \fi%
  \next%
}

\def\ric#1#2#3#4{%
  \def\next{\ric{#1}{#2}{#3}}%
  \ifx#4.%
     #3%
    \def\next{\relax}%
  \else\ifx#4|%
     \let\ricbox=\Fbox%
  \else\ifx#4?%
     \let\ricbox=\Ebox%
  \else%
    #2%
    \ricbox{#4}%
    \def\next{\ric{#1}{#1}{#3}}%
  \fi\fi\fi%
  \next%
}

\def\Ric#1#2#3#4{%
  \let\ricbox=\Ebox%
  \ric{#1}{#2}{#3}#4.%
}

\newbox{\first}
\newbox{\second}
\newbox{\third}
\newbox{\fourth}

\newdimen{\ringwd}
\newdimen{\ringht}

\def\Ring#1#2#3#4{%
  \setbox\first=\hbox{\Ric{\rightarrowfill}{}{}{#1}}%
  \setbox\second=\vbox{\Ric{\vdownarrowfill}{\vdownarrowfill}{\vdownarrowfill}{#2}}%
  \setbox\third=\hbox{\Ric{\leftarrowfill}{}{}{#3}}%
  \setbox\fourth=\vbox{\Ric{\vuparrowfill}{\vuparrowfill}{\vuparrowfill}{#4}}%
  \ringwd=\wd\first%
  \ifdim\ringwd<\wd\third%
    \ringwd=\ht\third%
  \fi%
  \ringht=\ht\second%
  \ifdim\ringht<\ht\fourth%
    \ringht=\ht\fourth%
  \fi%
  \advance\ringht by \ht\first%
  \advance\ringht by \ht\third%
  \vbox to \ringht{%
    \hbox to \ringwd {%
      \box\first%
    }%
    \hbox to \ringwd {%
          \box\fourth%
	  \hfill%
	  \box\second%
	  }%
    \hbox to \ringwd {%
	  \hfill%
      \box\third%
	  \hfill%
    }%
  }%
}

\ifx\PdfMode\empty
\DeclareGraphicsExtensions{.eps,.png,.ps,.jpg}
\else
\DeclareGraphicsExtensions{.pdf,.png,.ps,.eps,.jpg}
\fi

\def\Type/{{\sc Type}}
\def\Name/{{\sc Name}}
\def\Rate/{{\sc Rate}}
\def\inout/{\_\_inout\_\_}

\def\supp#1#2{#1}
\ifx\empty\empty
\def\lime/{{\em\supp{lime}{a-node}}}
\def\limes/{{\em\supp{limes}{a-nodes}}}
\def\HW/{\mac{HW}}\def\Hw/{\HW/}
\def\SW/{\mac{SW}}\def\Sw/{\SW/}
\def\SoD/{\supp{\mac{SoD}}{proprietary \mac{MPSoC} \mac{OS}}}
\else
\def\lime/{{\em\supp{actor}{a-node}}}
\def\limes/{{\em\supp{actors}{a-nodes}}}
\def\HW/{Hardware}\def\Hw/{hardware}
\def\SW/{Software}\def\Sw/{software}
\def\SoD/{\mac{OS}}
\fi

\def\slimer/{\cmd{\supp{slimer}{compiler}}}

\def\LIME/{\supp{\macs{LIME}}{Our \macs{PPM}}}
\def\Lime/{\supp{\macs{LIME}}{our \macs{PPM}}}
\def\NXP/{\supp{\macs{NXP}}{our company}}
\def\EVP/{\supp{\mac{EVP}}{{\em vector processing}}}
\def\MK/{\textsc{miniKanren}}
\def\Prolog/{\textsc{Prolog}}
\def\Boom/{\textsc{Boomerang}}
\def\Stratego/{\textsc{Stratego}}
\def\Maude/{\textsc{Maude}}

\def\mtom/{model$\mapsto$model}
\def\mtos/{model$\mapsto$source}

\def\nocite#1{}

\def\bul{\ensuremath{\bullet}}

\makeatletter
\newcommand{\xRightarrow}[2][]{\ext@arrow 0359\Rightarrowfill@{#1}{#2}}
\makeatother

\DeclareUnicodeCharacter{03B1}{\ensuremath{\alpha}}
\DeclareUnicodeCharacter{03B2}{\ensuremath{\beta}}
\DeclareUnicodeCharacter{03B8}{\ifmmode\theta\else{theta}\fi}
\DeclareUnicodeCharacter{00D7}{\ifmmode\times\else{*}\fi}
\DeclareUnicodeCharacter{00F7}{\ifmmode\div\else{/}\fi}
\DeclareUnicodeCharacter{02D6}{+}
\DeclareUnicodeCharacter{02D7}{-}

\DeclareUnicodeCharacter{03B5}{\ensuremath{\epsilon}}
\DeclareUnicodeCharacter{03C0}{\ensuremath{\pi}}
\DeclareUnicodeCharacter{03C9}{\ensuremath{\omega}}
\DeclareUnicodeCharacter{03B7}{\ensuremath{\eta}}
\DeclareUnicodeCharacter{03BA}{\ensuremath{\kappa}}
\DeclareUnicodeCharacter{00B2}{\ensuremath{^2}}
\DeclareUnicodeCharacter{1D6D1}{\ensuremath{\mathsf{\boldsymbol{\pi}}}}

\DeclareUnicodeCharacter{03BB}{\ensuremath{\lambda}}
\DeclareUnicodeCharacter{039B}{\ensuremath{\Lambda}}
\DeclareUnicodeCharacter{1D6CC}{\ensuremath{\mathsf{\lambda}}}
\DeclareUnicodeCharacter{1D77A}{\ensuremath{\mathsf{\boldsymbol{\lambda}}}}

\DeclareUnicodeCharacter{2758}{\vrule}
\DeclareUnicodeCharacter{00A6}{|}

\DeclareUnicodeCharacter{1D52}{\ensuremath{^\text{o}}}
\DeclareUnicodeCharacter{2261}{\ensuremath{\equiv}}

\DeclareUnicodeCharacter{2295}{\ensuremath{\oplus}}
\DeclareUnicodeCharacter{2297}{\ensuremath{\otimes}}

\DeclareUnicodeCharacter{21D0}{\ensuremath{\Leftarrow}}
\DeclareUnicodeCharacter{21D2}{\ensuremath{\Rightarrow}}
\DeclareUnicodeCharacter{2190}{\ensuremath{\leftarrow}}
\DeclareUnicodeCharacter{2192}{\ensuremath{\rightarrow}}
\DeclareUnicodeCharacter{2205}{\ensuremath{\emptyset}}
\DeclareUnicodeCharacter{2A04}{\ensuremath{\uplus}}
\DeclareUnicodeCharacter{22A5}{\ensuremath{\bot}}

\DeclareUnicodeCharacter{3008}{<}
\DeclareUnicodeCharacter{3009}{>}
\DeclareUnicodeCharacter{300A}{\guillemotleft}
\DeclareUnicodeCharacter{300B}{\guillemotright}

\DeclareUnicodeCharacter{00B7}{\ensuremath{\cdot}}
\DeclareUnicodeCharacter{266F}{\ensuremath{\sharp}}
\DeclareUnicodeCharacter{21A6}{\ensuremath{\mapsto}}

\DeclareUnicodeCharacter{2208}{\ensuremath{\in}}
\DeclareUnicodeCharacter{2209}{\ensuremath{\not\in}}

\DeclareUnicodeCharacter{00C1}{\'A}
\DeclareUnicodeCharacter{00E1}{\'a}

\DeclareUnicodeCharacter{00C9}{\'E}
\DeclareUnicodeCharacter{00E9}{\'e}
\DeclareUnicodeCharacter{00EF}{\"i}
\DeclareUnicodeCharacter{2026}{\ldots}
\DeclareUnicodeCharacter{22EF}{\cdots}
\DeclareUnicodeCharacter{2218}{\ensuremath{\circ}}
\DeclareUnicodeCharacter{22A4}{\ensuremath{\top}}
\DeclareUnicodeCharacter{22A5}{\ensuremath{\bot}}
\DeclareUnicodeCharacter{8596}{\ensuremath{\leftrightarrow}}
\DeclareUnicodeCharacter{2194}{\ensuremath{\leftrightarrow}}
\DeclareUnicodeCharacter{21D4}{\ensuremath{\Leftrightarrow}}
\DeclareUnicodeCharacter{207F}{\ensuremath{^n}}
\DeclareUnicodeCharacter{2070}{\ensuremath{^0}}

\def\ellipsis/{(…\hspace{-2pt})}

\def\nats{\ensuremath{\mathbb{N}}}
\def\ints{\ensuremath{\mathbb{Z}}}

\begin{document}
\maketitle


\begin{abstract}
I would like to tell a story. A story about a beautiful mathematical relationship that elucidates the computational view on the classic subject of trigonometry. All stories need a language, and for this particular story an \emph{algorithmic} language ought to do well. What makes a language algorithmic? From our perspective as the functional programming community, an algorithmic language provides means to express computation in terms of \emph{functions}, with no implementation-imposed limitations. We develop a new algorithm for the computation of trigonometric functions on dyadic rationals, together with the language used to express it, in Scheme. We provide a mechanically-derived algorithm for the computation of the inverses of our target functions. We address efficiency and accuracy concerns that pertain to the implementation of the proposed algorithm either in hardware or software.
\end{abstract}

\def\bintree/{\hspace{-3em}\raisebox{-20em}{\begin{tikzpicture}
[level 0/.style={level distance=2em},
 level 1/.style={level distance=2em,sibling distance=20.5em},
 level 2/.style={level distance=2em,sibling distance=10.3em},
 level 3/.style={level distance=2em,sibling distance=5.25em},
 level 4/.style={level distance=2em,sibling distance=2.56em},
]
\node{\bul} [edge from parent fork left,grow=left]
child{node{\bul}
 child {node{\bul}
  child {node{\bul}
 }
  child {node{\bul}
 }
 }
 child {node{\bul}
  child {node{\bul}
 }
  child {node{\bul}
 }
 }
}
child{node{\bul}
 child {node{\bul}
  child {node{\bul}
 }
  child {node{\bul}
 }
 }
 child {node{\bul}
  child {node{\bul}
 }
  child {node{\bul}
 }
 }
};
\end{tikzpicture}}}


\section{Introduction}
\label{Introduction}

The use of square roots and circular (trigonometric) functions is pervasive in computer science and engineering. In computational geometry and physics, for example, both square roots and trigonometric functions are used extensively. In many engineering domains, the evaluation of both trigonometric functions and their inverses is required for e.g., pre-computation of the \mac{FFT} tables, angle calculations and rotations of complex numbers. For digital signal processing in general and in wireless communications systems in particular, efficient and accurate implementation of such functions is fundamental (e.g., for frequency error estimation and correction).

In this paper we argue that square roots and their infinite continuation (hereafter: nested radicals) deserve to be quite a bit more pervasive in computing, just the more so because of the pervasiveness of dyadic rationals. These are the numbers of the form $x={n\over m}, m=2^k, n\in\ints, k\in\nats$. Both embedded software and hardware are generally designed using a fixed-point representation of (complex) values. In such a representation of a number, a fixed number of bits ($W$) are allocated for both the \emph{integer part} ($I\ge 0$, which may include the sign bit), leaving the rest of the bits for the \emph{fractional} part ($F=W-I$ bits). This naturally corresponds to dyadics ${n\over 2^F}, n\in[0,2^W-1]$ (or $[-2^{W-1},2^{W-1}-1]$ for two's complement representation of negative numbers). Such fixed-point types are very frequently used in embedded systems as they possess predictable precision and resource efficiency characteristics. Conversion to a dyadic rational is a simple division by $2^F$ followed by rational simplification.

And yet, maintaining explicit tables containing fixed-point values for e.g., \mac{FFT} or a \mac{CORDIC} \cite{volder2000birth} can be too expensive for resource-constrained implementations. For the software, the efficiency is limited by the memory hierarchy bottleneck while for hardware excessive (de)multiplexing that is associated to large (frequently $>4Kbit$) tables can outweigh the advantages of avoiding explicit multipliers - the primary reason for tabling as opposed to Taylor/Maclaurin series. There is a trade-off between multiplier-based approach with series and multiplexer-based approach with tabling that a resource-constrained application may not want to make.

Maybe we can get rid of explicit tables of values inside implementations of trigonometric functions at the expense of restricting the domain to dyadic rationals? We set our goal therefore to provide a computational proof of this conjecture, by presenting an algorithm for the evaluation of trigonometric functions and their inverses. We shall be eliding tables and/or Taylor/Maclaurin series, using only the very basic arithmetic operations that can map well to hardware and embedded software architectures. These are: addition/subtraction, multiplication/division by a small power-of-2 (bit-shifts), squaring, square root and reciprocals. Note that all these operations except the last three map trivially to modern digital logic. Squaring may be performed without full multipliers using one of the existing methods \cite{Sethi2015} and \cite{deshpande2010squaring}, while the square root may be computed using a fixed-bound iteration described below.

This paper is structured as follows. First, in section \ref{Development}, we will focus on the development of our \emph{reversible} algorithm starting from the basics of trigonometry, as well as the basics of \mac{R5RS}. We use the \emph{Algorithmic} Language Scheme \cite{R5RS} primarily because of its flexible syntax and conceptual simplicity. Scheme, to our knowledge, is the only language besides Algol and Refal (both of which we shall avoid here) specifically targeting ``algorithms''. We will, however, abstain from the impure subset of Scheme, making the code directly translatable to Haskell, \mac{ML} or indeed any other functional language. We shall incidentally use the simple algorithm for the square root to highlight our choice for Scheme as an implementation language.

Gradually, we will introduce a number of extensions to both the algorithm and the language used to implement it. Scheme-related sections are tagged as ``tangential'' to highlight their supplementary nature. Our final result is evaluated and compared to prior art in section \ref{Evaluation}. We conclude and cast the light on future work in section \ref{Conclusions}.

\section{Development of the algorithm}
\label{Development}

In this section we present observations to show that $\cos{x\over 2^k}$ can indeed be computed iteratively using nested radicals, provided that $x$ is an even multiple of ${\pi\over 4}$ (i.e., a multiple of ${\pi\over 2}$). For the rest of the paper, we will be restricting the domain of the argument ${x\over 2^k}\in[0,{\pi\over 2}]$ as well as factoring it by ${\pi\over 2}$, thus keeping only integers $n\in[0,2^k], n={2x\over\pi}$ such that the argument is always of the form ${n\pi\over 2m}, m=2^k$. This effectively partitions the range into an exponential number of equally-spaced bins forming a \emph{lattice}, where each bin is represented by a dyadic rational ${n\over m}$. For each bin, our focus is to find an efficient and accurate method to compute $\cos{n\pi\over 2m}, m=2^k, n\in\ints, k\in\nats$.

It is not difficult to see why nested radicals appear naturally in evaluation of circular functions (also called trigonometric functions) such as $\cos{}$\&$\sin{}$ on modern hardware. One may remember from the trigonometry lessons that squares of the (results of) these two functions add up to 1. A complex number using polar representation (comprising magnitude and angle) naturally corresponds to the conjoined (results of) trigonometric functions acting on the angle, scaled by the magnitude. I.e., projections of a complex number onto a rectangular, or Cartesian coordinate system follows the rules of trigonometry, and therefore simple geometric observations on a two dimensional plane (such as the Pythagoras's theorem) can also serve as a good illustration of the trigonometric facts that are important in this paper:

\begin{enumerate}

\item There is an angle, ${\pi\over 4}$ for which both projections are equal: the \emph{Pythagorean identity} implies that $\sin{\pi\over 4}=\cos{\pi\over 4}={1\over\sqrt{2}}=\sqrt{1\over 2}$.

\item For any two angles, the addition of angles can be performed by a multiplication of the corresponding complex numbers having magnitude 1. This can be used to show how to compute trigonometric functions of the argument which is a double (or indeed, any multiple) of an angle, given (results of) trigonometric functions of that angle. Similarly, the \emph{half-angle} formul\ae{} tell us how to reduce the computation of e.g., $\cos{x\over 2}$ to $(-1)^{\lfloor{x+\pi\over 2\pi}\rfloor}\sqrt{1+\cos{x}\over 2}$.

\item Combining above two observations, we get the following sequence:

\noindent $cos{x\over 2^k}=s^k_0(x)\sqrt{1+\cos{x\over 2^{k-1}}\over2}=s^k_0(x)\sqrt{{1\over 2}+s^k_1(x)\sqrt{1+\cos{x\over 2^{k-2}}\over 2^2*2}}=s^k_0(x)\sqrt{{1\over 2^0*2}+s^k_1(x)\sqrt{{1\over 2^2*2}+s^k_2(x)\sqrt{...}}}$

\noindent $=s^k_0(x)\sqrt{{2\over 2^{2^1}}+s^k_1(x)\sqrt{{2\over 2^{2^2}}+s^k_2(x)\sqrt{{2\over 2^{2^3}}+...}}}$ where $s^k_i(x)=(-1)^{\lfloor{{x\over \pi{}2^{k-i}}+{1\over 2}}\rfloor}$.

\end{enumerate}

One natural question to ask is whether the above nested, or \emph{continued} radical is actually well-founded. We see that the repeated doubling of the argument with subsequent reversal of the sign $s^k_i(x)$ whenever ${x\over 2^k}>{\pi\over 2}$ (where $\cos{x}$ becomes negative) leads to an oscillation of the value of our target function $\cos{}$ around $0$, with smaller and smaller radical terms. Restricting precision of the term to match desired accuracy, we will inevitably hit $\cos{\pi\over 2}=0$, terminating the recursion. A number of signs of $s^4_i(n)$ for $n\in[1,15]$ and $i\in[1,3]$ (when defined, depending on $n$ as explained in section \ref{First target: Cosine}) is given in Figure \ref{dy16}.

\ifx0
Both restrictions are soft: the former can be lifted by making use of the trigonometric symmetries (addressed in sections \ref{First target: Cosine} and \ref{Evaluation}) while the latter can be lifted by increasing the precision of the terms. The algorithm itself will not require modifications in either case.
\fi

So we can indeed compute $\cos{n\pi\over 2m}$ using iterated \texttt{sqrt} computation. We can compute the sign sequence $s^k_i(n)$ which is uniquely determined by the index of $n$ as the dyadic rational in the lattice induced by $2^{-k}$, and we know that the process stops as soon as the next positive radical term can no longer be represented. But how do we compute the \texttt{sqrt} in the first place?

\subsection{The (square) root of all good
}
\label{The (square) root of all good}

Of all known methods to compute square roots \cite{nikulin2001hazewinkel} the following suits our needs the best. We start with an estimation of the $error^2=2^{2i}$ (the highest power of 4 that is smaller than the argument), and an estimation of the $result=0$. Then we gradually re-balance the equation $number=(result+error)^2=result^2+2\times result\times error+error^2$ by adding bits to the $result$ (focusing on the second term) so that the $error$ is minimised. A popular {\CEE/} implementation \cite{Martin} cites \cite{Woo} as the source of this method.

{\small{\begin{flushleft}
{
{\let\oldtbf=\textbf\def\textbf#1{{\rm\oldtbf{#1}}}\noindent \texttt{\begin{tabular*}{0.90\linewidth}{l@{\extracolsep{\fill}}}
(\textbf{define}\ (sqrt-loop\ n\ start)\\
\ \ \ (\textbf{let}\ loop\ ({\char91}number\ n{\char93}\\
\ \ \ \ \ \ \ \ \ \ \ \ \ {\char91}result\ 0{\char93}\\
\ \ \ \ \ \ \ \ \ \ \ \ \ {\char91}error\ start{\char93})\\
\ \ \ \ \ \ (\textbf{cond}\\
\ \ \ \ \ \ \ \ \ \ \ \ ({\char91}zerofx?\ error{\char93}\ result)\\
\ \ \ \ \ \ \ \ \ \ \ \ ({\char91}>=fx\ number\ (+fx\ result\ error){\char93}\\
\ \ \ \ \ \ \ \ \ \ \ \ \ (loop\ (-fx\ number\ (+fx\ result\ error))\\
\ \ \ \ \ \ \ \ \ \ \ \ \ \ \ \ (+fx\ error\ (bit-rsh\ result\ 1))\\
\ \ \ \ \ \ \ \ \ \ \ \ \ \ \ \ (bit-rsh\ error\ 2)))\\
\ \ \ \ \ \ \ \ \ \ \ \ (\textbf{else}\\
\ \ \ \ \ \ \ \ \ \ \ \ \ (loop\ number\\
\ \ \ \ \ \ \ \ \ \ \ \ \ \ \ \ (bit-rsh\ result\ 1)\\
\ \ \ \ \ \ \ \ \ \ \ \ \ \ \ \ (bit-rsh\ error\ 2)))\\\ \ \ )))\\
\end{tabular*}
}}
}\end{flushleft}
}}

All this can be accomplished \emph{tail-recursively} in Scheme with efficient right bit-shifts (that correspond to repeated halving) and additions on fixnums (which in our dialect of Scheme (Bigloo) are invoked using ``fx'' suffix). The \texttt{loop} binding captures the body of the iteration, which is parameterised by the ``variables'' involved in the re-balancing. Initial values are given at the first invocation (inside \texttt{let}), and the \texttt{cond} clauses specify how these should be ``updated'' in each case, when the \texttt{loop} is invoked (arguments are positional, following the order given by the \texttt{let}). Because in each iteration exactly 2 bits are right-shifted out from the $error$, the iteration limit linearly depends on the precision of used fixnums. This value will eventually reach 0, at that time we can return the current approximation of the $result$.

{\small{\begin{flushleft}
{
{\let\oldtbf=\textbf\def\textbf#1{{\rm\oldtbf{#1}}}\noindent \texttt{\begin{tabular*}{0.90\linewidth}{l@{\extracolsep{\fill}}}
(\textbf{def-syntax}\ *maxbit*\ 60)\\
(\textbf{define}\ (sqrt-start\ n)\\
\ (\textbf{do}\ ({\char91}bit\ (bit-lsh\ 1\ (-fx\ *maxbit*\ 2))\ (bit-rsh\ bit\ 2){\char93})\\
\ \ \ \ \ ({\char91}<=fx\ bit\ n{\char93}\ bit)\\
\ \ \ ))\\
\end{tabular*}
}}
}\end{flushleft}
}}

Because we may already know highest power of 4 before the \texttt{sqrt} loop is run, we pass a parameter (\texttt{start}) to be computed separately beforehand, if needed. E.g., by the \texttt{sqrt-start} function above.
Here, we assume a maximal precision for the fixnums, specified as a syntactic ``constant'' \texttt{*maxbit*}. We search for the largest power of 4 that is smaller than \texttt{n} using Scheme's iterator construct, that macro-expands to a tail-recursion similar to the one we used in \texttt{sqrt-loop}. As we have seen above, the calculation of trigonometric functions also involves even powers of 2, so adding this argument might avoid extraneous invocation of \texttt{sqrt-start} if the value of (\texttt{n}) is bounded by a known integer.

\begin{figure}[htb]
\label{Values from the hexadecaic interval}
\center\begin{tabular}{cccccc}
\textbf{\textsf{$s^k_i(n)$ }} & \textbf{\textsf{$n\over m$          }} & \textbf{\textsf{$\cos{{n\pi\over 2m}}$                                        }} & \textbf{\textsf{\texttt{sig}     }} & \textbf{\textsf{\texttt{sig'}      }} & \textbf{\textsf{tree}} \\\hline\\
     (+,+,+)  & ${1\over 16}$ &  $\sqrt{{2^{-1}}+\sqrt{{2^{-3}}+\sqrt{{2^{-7}}+\sqrt{2^{-15}}}}}$ & 0            & 16           \\
       (+,+)  & ${1\over 8}$  &  $\sqrt{{2^{-1}}+\sqrt{{2^{-3}}+\sqrt{2^{-7}}}}$                & 0            & 8            \\
     (+,+,-)  & ${3\over 16}$ &  $\sqrt{{2^{-1}}+\sqrt{{2^{-3}}+\sqrt{{2^{-7}}-\sqrt{2^{-15}}}}}$ & 1            & 17           \\
         (+)  & ${1\over 4}$  &  $\sqrt{{2^{-1}}+\sqrt{2^{-3}}}$                                & 0            & 4            \\
     (+,-,-)  & ${5\over 16}$ &  $\sqrt{{2^{-1}}+\sqrt{{2^{-3}}-\sqrt{{2^{-7}}-\sqrt{2^{-15}}}}}$ & 3            & 19           \\
       (+,-)  & ${3\over 8}$  &  $\sqrt{{2^{-1}}+\sqrt{{2^{-3}}-\sqrt{2^{-7}}}}$                & 1            & 9            \\
     (+,-,+)  & ${7\over 16}$ &  $\sqrt{{2^{-1}}+\sqrt{{2^{-3}}-\sqrt{{2^{-7}}+\sqrt{2^{-15}}}}}$ & 2            & 18           \\
          ()  & ${1\over 2}$  &  $\sqrt{2^{-1}}$                                                & 0            & 2            \\
     (-,-,+)  & ${9\over 16}$ &  $\sqrt{{2^{-1}}-\sqrt{{2^{-3}}-\sqrt{{2^{-7}}+\sqrt{2^{-15}}}}}$ & 6            & 22           \\
       (-,-)  & ${5\over 8}$  &  $\sqrt{{2^{-1}}-\sqrt{{2^{-3}}-\sqrt{2^{-7}}}}$                & 3            & 11           \\
     (-,-,-)  & ${11\over 16}$  &  $\sqrt{{2^{-1}}-\sqrt{{2^{-3}}-\sqrt{{2^{-7}}-\sqrt{2^{-15}}}}}$ & 7            & 23           \\
         (-)  & ${3\over 4}$  &  $\sqrt{{2^{-1}}-\sqrt{2^{-3}}}$                                & 1            & 5            \\
     (-,+,-)  & ${13\over 16}$  &  $\sqrt{{2^{-1}}-\sqrt{{2^{-3}}+\sqrt{{2^{-7}}-\sqrt{2^{-15}}}}}$ & 5            & 21           \\
       (-,+)  & ${7\over 8}$  &  $\sqrt{{2^{-1}}-\sqrt{{2^{-3}}+\sqrt{2^{-7}}}}$                & 2            & 10           \\
     (-,+,+)  & ${15\over 16}$  &  $\sqrt{{2^{-1}}-\sqrt{{2^{-3}}+\sqrt{{2^{-7}}+\sqrt{2^{-15}}}}}$ & 4            & 20           \\\hline
\end{tabular}
\scalebox{.8}{\bintree/}
\caption{{\color{black}\label{dy16}
}{\color{black}}Values from the hexadecaic interval}
\end{figure}

We could stop here - the algorithm is simple enough. The purists would of course object and complain about the ugliness of the code, the abundance of fixed-point modifiers, parentheses, etc. Therefore we will \emph{not} stop, and proceed to actually extend Scheme with \emph{anatypes} (short for \emph{anaphoric} types), described in section \ref{Tangential: anatypes} which is included as supplementary material since this largely concerns only syntactic improvements. All such syntactic improvements (implemented as macros) will be used for code snippets in the followup. Additionally, from now on, most binding forms will be capitalised to highlight the fact that the language they use is Scheme extended with anatype annotations as in e.g., Figure \ref{cos1}.

\subsection{First target: Cosine
}
\label{First target: Cosine}

Returning to our goal of implementing all trigonometric functions, let us first focus on the simplest one: the cosine. One may remember from elementary trigonometry that in order to compute $\cos{x}$ it is only needed to find a value for $x\in[0,{\pi\over 2}]$, the rest can be found by symmetry (please see section \ref{Evaluation} for a common way to resolve this). Also, we only need to implement just one of these functions really, since standard trigonometric identities can be used to derive all the other ones.

So in our quest let us focus on just the values of $\cos{n\pi\over 2m}$ for dyadic rationals ${n\over m}$ within this interval. Remember that we are re-scaling the inputs to our function by ${\pi\over 2}$, keeping dyadics ${n*2^{-k}}, 0\le{n}\le{2^k}$. Leaving out ${\pi\over 2}$ as an implicit domain/range factor will also simplify reverse-mode trigonometric functions, making them more precise at the same time (see section \ref{Inverse trigonometric functions}).
A first observation is that $\cos{0}=1$ or $\cos{2^k\pi\over 2*2^k}=0$ so the exact value is already known. A second observation is that this function is strictly \emph{monotonically decreasing}. This means that when the argument $n*2^{-k}\rightarrow 0$ the result should approach 1 from below. So for example for all rationals ${1\over 2}, {1\over 4} ... {1\over 2^k}$ the signs are necessarily positive: $s^k_i(n)=1$ (as was already shown by \cite{zimmerman2008infinitely}), since the \texttt{sqrt} function itself is strictly monotonically \emph{increasing}.

But how do we compute correct signs for other dyadic rationals in the wanted domain? Immediate use of the definition of $s^k_i(n)$ seems prohibitive from the efficiency perspective. Looking at a few values in Figure \ref{dy16} we can observe the pattern: as the numerator increases, the signs change from positive, then to zero (indicated by the absence of a sign in the left column) and then to negative (in the first half). If there is an odd number of negative signs to the left inside the nested radical (in the sequence $s^k_j(n), j<i$), then the order reverses to first negative, then zero and then positive. One can also observe from Figure \ref{dy16} that for the bottom half of the domain, the order of signs is reversed w.r.t. the top half.
So we can consider the evaluation of such functions as finding a ``signature'', or a position, of each dyadic rational in a \emph{binary search} ``tree'' (depicted in the rightmost column of Figure \ref{dy16}), determined by the linear partitioning of the sector into an exponential number of equally-spaced sub-sectors (minus one, since we are excluding $0$ and $1$). Because of the monotonicity of the target function ($\cos{x}$) and the function used to implement it ($\sqrt{x}$), the signature must exhibit a memory-like behaviour, recording the number of negative signs in the nested radical expansion. Once a single (or an odd number of) negative sign(s) is reached, all further signs are inverted. An involution (i.e., a multiple of two of negative turns) makes the sign positive again.

To implement the signature we can use a simple bit-mask \texttt{sig} (column 4 in Figure \ref{dy16}) that contains the bit 1 in the position of each negative turn. Since a \emph{continued} radical is best defined in a continued form, an obvious approach in functional programming tradition for implementing an approximation to the continued radical above is to use the venerable \emph{continuation passing style}. In other words, on every expansion (i.e., at each level of the binary search) of the half-angle formula, we take the following steps:

\begin{enumerate}

\item compute the term ${2\over 2^{2^i}},i\ge 1$ as \texttt{(* 2 f)}
\begin{enumerate}

\item starting from a rational $f={1\over 4}$ or flonum \texttt{(exp2 -2)}

\item for each level ($i$), repeatedly square it

\end{enumerate}

\item add the continuant to the term

\item take the square root, in either fixnums or flonums

\item compute the sign using bit-mask \texttt{sig} (masking the last turn, bit 0)

\item deliver the signed result to the continuation

\end{enumerate}

\begin{figure}[htb]
\label{Computing dyadic cosine using continuations}
\fbox{\hbox to .99\columnwidth{\vbox{\vspace{-.5em}{\small{\begin{flushleft}
{
{\let\oldtbf=\textbf\def\textbf#1{{\rm\oldtbf{#1}}}\noindent \texttt{\begin{tabular*}{0.90\linewidth}{l@{\extracolsep{\fill}}}
(\textbf{Define}\ (dyadic-cos1\ n\ \textmd{\textit{:fix}}\ m\ \textmd{\textit{:fix}}\ sig\ \textmd{\textit{:fix}}\ f\ \textmd{\textit{:flo}}\ cont\ \textmd{\textit{:proc}})\\
\ (\textbf{cond}\ ({\char91}\textbf{=}\ n\ 0{\char93}\ (cont\ 1))\ \ \ \ \ \ \ \ \ \ \ \ \ \ \ \ \ \ \ \ \ \ \ \ \ \ \ \ \ \ \ \ \ \ \ \ \ \ \ \textit{;; exact }\\
\ \ \ \ \ \ \ ({\char91}\textbf{=}\ n\ m{\char93}\ (cont\ 0))\ \ \ \ \ \ \ \ \ \ \ \ \ \ \ \ \ \ \ \ \ \ \ \ \ \ \ \ \ \ \ \ \ \ \ \ \ \ \ \textit{;; results}\\
\ (\textbf{else}\ (\textbf{Let*}\ (fix:{\char91}m\ (\textmd{\textit{rsh}}\ m\ 1){\char93}\\
\ \ \ \ \ \ \ \ \ \ \ \ \ \ \ \ \ {\char91}sign\ (\textbf{if}\ {\char91}\textbf{=}\ (\textit{bit-and}\ sig\ 1)\ 1{\char93}\ -1\ 1){\char93})\\
\ \ (\textbf{cond}\ ({\char91}\textbf{=}\ n\ m{\char93}\\
\ \ \ \ \ \ \ \ \ (cont\ (\textbf{*}\ sign\ (\textbf{sqrt}\ (\textbf{*}\ 2\ f)))))\\
\ \ \ \ \ \ \ \ ({\char91}<\ n\ m{\char93}\\
\ \ \ \ \ \ \ \ \ (dyadic-cos1\ n\ m\ \\
\ \ \ \ \ \ \ \ \ \ \ \ (\textmd{\textit{bit-or}}\ (\textmd{\textit{bit-lsh}}\ sig\ 1)\\
\ \ \ \ \ \ \ \ \ \ \ \ \ \ \ (\textbf{if}\ (\textmd{\textit{even?}}\ (\textmd{\textit{bit-count}}\ sig))\\
\ \ \ \ \ \ \ \ \ \ \ \ \ \ \ \ \ \ \ 0\ 1))\\
\ \ \ \ \ \ \ \ \ \ \ \ (\textbf{sqr}\ f)\\
\ \ \ \ \ \ \ \ \ \ \ \ (λ\ (x)\ (cont\ (\textbf{*}\ sign\ (\textbf{sqrt}\ (\textbf{+}\ (\textbf{*}\ 2\ f)\ x)))))\\
\ \ \ \ \ \ \ \ \ \ \ \ ))\\
\ \ \ \ \ \ \ \ ({\char91}>\ n\ m{\char93}\\
\ \ \ \ \ \ \ \ \ (dyadic-cos1\ (\textbf{-}\ n\ m)\ m\ \\
\ \ \ \ \ \ \ \ \ \ \ \ (\textmd{\textit{bit-or}}\ (\textmd{\textit{bit-lsh}}\ sig\ 1)\ \\
\ \ \ \ \ \ \ \ \ \ \ \ \ \ \ (\textbf{if}\ (\textmd{\textit{even?}}\ (\textmd{\textit{bit-count}}\ sig))\\
\ \ \ \ \ \ \ \ \ \ \ \ \ \ \ \ \ \ \ 1\ 0))\\
\ \ \ \ \ \ \ \ \ \ \ \ (\textbf{sqr}\ f)\\
\ \ \ \ \ \ \ \ \ \ \ \ (λ\ (x)\ (cont\ (\textbf{*}\ sign\ (\textbf{sqrt}\ (\textbf{+}\ (\textbf{*}\ 2\ f)\ x)))))\\
\ \ \ \ \ \ \ \ \ \ \ \ )))))))\\
\end{tabular*}
}}
}\end{flushleft}
}}\vspace{-.5em}}}}
\vspace{-2em}\caption{{\color{black}\label{cos1}
}{\color{black}}Computing dyadic cosine using continuations}
\end{figure}

In Figure \ref{cos1} we see that as soon as bisection of the interval hits the rational ${n\over m}$ corresponding to the argument of $\cos{n\pi\over 2m}$, the continuant can be assumed to be 0 and the value can finally be delivered to the chain of continuations that we have built by this tail-recursion. On every level, the interval is halved (\texttt{m} and \texttt{n} steadily approach each other, the former by halving and the latter by subtraction whenever $n>m$). The signature is maintained using bit-shift and bit-counting according to the scheme described above. Please see Figure \ref{dy16} for successive values of \texttt{sig}, which appear to behave quite erratically. However, it can be shown that \texttt{sig}, in fact, is an example of an infinite \emph{fractal} sequence (containing itself as a sub-sequence) when one concatenates sequences for each successive lattice $2^{-k}, k=[1,2,3 ...]$: \texttt{{\char91}0{\char93} {\char91}0 0 1{\char93} {\char91}0 0 1 0 3 1 2{\char93} {\char91}0 0 1 0 3 1 2 0 6 3 7 1 5 2 4{\char93}...}.
We have attempted to derive an explicit formula for computing \texttt{sig}, by priming it with one extra bit in front (and reversing bit-counting logic in the algorithm of course, see section \ref{Control and representation optimisations}). Resulting values of \texttt{sig'} (column 5 in Figure \ref{dy16}) look regular, however, the pattern depends on both \texttt{n} and \texttt{m} and exhibits fractal mirroring, so further simplifications remain elusive (fractal sequences from the OEIS A101279 for binary trees, A131987 for dyadic rationals and A025480 for the Grundy values do appear to be related to our sequence for \texttt{sig'}).

Running \texttt{(dyadic-cos1 n m {\char35}b0 (/ 1 4) ($\lambda$ (x) x))} is an effective way to compute $\cos{n\pi\over 2m}$ and costs $O(\log{m})$ evaluations of \texttt{sqrt} at each level (all logarithms in this paper are base-2) requiring constant time each. No tables are in sight and the result is available at varying levels of accuracy. The fixed-point operator modifiers are replaced by anatype annotations for each binding: the inputs integers use \texttt{fix:} while the output floats use \texttt{flo:}, in either case, appropriate type-specialised arithmetic operators are automatically substituted in place of their generic versions. Below, we shall also elide prefixes from bitwise operators e.g., $\textmd{\textit{bit-and}}\mapsto\textit{and}$, typesetting logical operators such as $\textmd{\textbf{and}}$ to distinguish.

\subsection{Following Reynolds ...
}
\label{Following Reynolds ...}

Let us proceed in improving our algorithm for $\cos{n\pi\over 2m}$. The code in Figure \ref{cos1} contains many nested expressions for computing various quantities. It would be nice if we could \emph{thread} the temporary values through such expressions without having to deal with deep nesting that is inevitable with a language such as Scheme. In section \ref{Tangential: threading} we introduce our macros for threading which help to alleviate the problem. The \texttt{>->} form used in code below takes a flat list of expressions (which can be read inside-out, from last to first) and substitutes in each expression all occurrences of \texttt{\_} (i.e., \emph{anaphroric} references) by the syntax of the preceding expression. Again, this is another straightforward syntactic transformation that will prove helpful in later code snippets in this paper.

\begin{figure}[htb]
\fbox{\hbox to .99\columnwidth{\vbox{\vspace{-.5em}{\small{\begin{flushleft}
{
{\let\oldtbf=\textbf\def\textbf#1{{\rm\oldtbf{#1}}}\noindent \texttt{\begin{tabular*}{0.90\linewidth}{l@{\extracolsep{\fill}}}
(\textbf{Define}\ (dyadic-cos2\ fix:\ n\ m\ sig\ f\ lf)\\
(\textbf{cond}\ ({\char91}\textbf{=}\ n\ 0{\char93}\ 1)\ \ \ \ \ \ \ \ \ \ \ \ \ \ \ \ \ \ \ \ \ \ \ \ \ \ \ \ \ \ \ \textit{;; exact }\\
\ \ \ \ \ \ ({\char91}\textbf{=}\ n\ m{\char93}\ 0)\ \ \ \ \ \ \ \ \ \ \ \ \ \ \ \ \ \ \ \ \ \ \ \ \ \ \ \ \ \ \ \textit{;; results}\\
(\textbf{else}\ (\textbf{Let*}\ ({\char91}m\ \textmd{\textit{:fix}}\ (\textmd{\textit{rsh}}\ m\ 1){\char93})\\
\ (\textbf{cond}\ ({\char91}\textbf{=}\ n\ m{\char93}\ (apply2\ sig\ f\ lf))\\
\ \ \ \ \ \ \ ({\char91}<\ n\ m{\char93}\ (dyadic-cos2\ n\ m\\
\ \ \ \ \ \ \ \ \ \ \ \ \ \ \ \ \ \ \ (>->\ sig\ (\textmd{\textit{lsh}}\ \_\ 1)\\
\ \ \ \ \ \ \ \ \ \ \ \ \ \ \ \ \ \ \ \ \ \ (\textit{or}\ \_\ (\textbf{if}\ (\textmd{\textit{even?}}\ (\textmd{\textit{bit-count}}\ sig))\\
\ \ \ \ \ \ \ \ \ \ \ \ \ \ \ \ \ \ \ \ \ \ \ \ \ \ \ \ \ \ \ \ 0\ 1)))\\
\ \ \ \ \ \ \ \ \ \ \ \ \ \ \ \ \ \ \ (\textmd{\textit{lsh}}\ f\ lf)\\
\ \ \ \ \ \ \ \ \ \ \ \ \ \ \ \ \ \ \ (\textmd{\textit{lsh}}\ lf\ 1)\\
\ \ \ \ \ \ \ \ \ \ \ \ \ \ \ \ \ \ \ ))\\
\ \ \ \ \ \ \ ({\char91}>\ n\ m{\char93}\ (dyadic-cos2\ (\textbf{-}\ n\ m)\ m\\
\ \ \ \ \ \ \ \ \ \ \ \ \ \ \ \ \ \ \ (>->\ sig\ (\textmd{\textit{lsh}}\ \_\ 1)\ \\
\ \ \ \ \ \ \ \ \ \ \ \ \ \ \ \ \ \ \ \ \ \ (\textit{or}\ \_\ (\textbf{if}\ (\textmd{\textit{even?}}\ (\textmd{\textit{bit-count}}\ sig))\\
\ \ \ \ \ \ \ \ \ \ \ \ \ \ \ \ \ \ \ \ \ \ \ \ \ \ \ \ \ \ \ \ 1\ 0)))\\
\ \ \ \ \ \ \ \ \ \ \ \ \ \ \ \ \ \ \ (\textmd{\textit{lsh}}\ f\ lf)\\
\ \ \ \ \ \ \ \ \ \ \ \ \ \ \ \ \ \ \ (\textmd{\textit{lsh}}\ lf\ 1)\\
\ \ \ \ \ \ \ \ \ \ \ \ \ \ \ \ \ \ \ )))))))\\
\end{tabular*}
}}
}\end{flushleft}
}}\vspace{-.5em}}}}
\label{Defunctionalized dyadic cosine}
\vspace{-2em}\caption{{\color{black}\label{cos2}
}{\color{black}}Defunctionalized dyadic cosine}
\end{figure}

Introduction of a bit-mask \texttt{sig} tracking the number of negative signs as the algorithm dichotomises the interval $[0,{\pi\over 2}]$ has a benevolent side-effect: it provides enough information to actually recreate the path to the wanted value in a corresponding binary search ``tree''. This can be captured by \emph{defunctionalized} \cite{Reynolds} version of the algorithm found in Figure \ref{cos2}. With defunctionalization of a higher-order function such as \texttt{dyadic-cos1}, we are constructing an object (i.e., a data-structure according to tradition in defunctionalization literature) instead of a chain of continuations built with $\lambda$s. This object is interpreted in a separate function that re-creates all the needed steps. In this particular case, the object is just the integer \texttt{sig}, which is unwound by right bit-shifts found in Figure \ref{apply2}.

Our new algorithm also uses bit-shifts to avoid squaring, using a new fixnum \texttt{f} and its $\log{f}=lf$ (initial $f_0=4, lf_0=2$). Each time, \texttt{f} is ``squared'' by multiplying it with $2^{lf}$, while \texttt{lf} is multiplied by 2. The bit-mask \texttt{sig} is used in a separate function, \texttt{apply2} (see Figure \ref{apply2}) where it is replayed to recreate all the turns of the binary search.
Because now factor \texttt{f} is a fixnum value, a conversion between \texttt{fix:} and \texttt{flo:} anatypes is needed (expanding to standard \texttt{exact->inexact} library call). Note that now we can also play with other defunctionalizations. For example, we can replace \texttt{(/ 2 (toflo f)}  by \texttt{(exp2 (- 1 lf))}, keeping only \texttt{lf} in the loop (elided here).

\begin{figure}[htb]
\label{Apply for defunctionalized dyadic cosine}
\fbox{\hbox to .99\columnwidth{\vbox{\vspace{-.5em}{\small{\begin{flushleft}
{
{\let\oldtbf=\textbf\def\textbf#1{{\rm\oldtbf{#1}}}\noindent \texttt{\begin{tabular*}{0.90\linewidth}{l@{\extracolsep{\fill}}}
(\textbf{Define}\ (apply2\ fix:\ sig\ f\ lf)\\
\ (\textbf{Let}\ loop\ (fix:{\char91}s\ sig{\char93}\\
\ \ \ \ \ \ \ \ \ \ \ \ \ \ \ \ {\char91}f\ f{\char93}\\
\ \ \ \ \ \ \ \ \ \ \ \ \ \ \ \ {\char91}lf\ (\textmd{\textit{rsh}}\ lf\ 1){\char93}\ \\
\ \ \ \ \ \ \ \ \ \ \ \ \ \ \ \ {\char91}a\ \textmd{\textit{:flo}}\ 0{\char93})\\
\ \ \ \ (\textbf{if}\ (\textmd{\textit{zero?}}\ lf)\ \\
\ \ \ \ \ \ \ \ a\\
\ \ \ \ \ \ \ \ (loop\ (\textmd{\textit{rsh}}\ s\ 1)\\
\ \ \ \ \ \ \ \ \ \ \ (\textmd{\textit{rsh}}\ f\ lf)\\
\ \ \ \ \ \ \ \ \ \ \ (\textmd{\textit{rsh}}\ lf\ 1)\\
\ \ \ \ \ \ \ \ \ \ \ (\textbf{*}\ (\textbf{if}\ {\char91}\textbf{=}\ (\textit{and}\ s\ 1)\ 1{\char93}\ -1\ 1)\\
\ \ \ \ \ \ \ \ \ \ \ \ \ \ (\textbf{sqrt}\ (\textbf{+}\ (\textbf{/}\ 2\ (toflo\ f))\\
\ \ \ \ \ \ \ \ \ \ \ \ \ \ \ \ \ \ \ \ \ \ \ a)))))))\\
\end{tabular*}
}}
}\end{flushleft}
}}\vspace{-.5em}}}}
\vspace{-2em}\caption{{\color{black}\label{apply2}
}{\color{black}}Apply for defunctionalized dyadic cosine}
\end{figure}

\subsection{Control and representation optimisations
}
\label{Control and representation optimisations}

The fact that using our recursion approach, vanishingly small radical terms need to be evaluated (e.g., \texttt{(sqr f)}, \texttt{(/ 2 (toflo f))} or \texttt{(exp2 (- 1 lf))}) means that the accuracy of our result (evaluation of a trigonometric function) is limited by the precision of the minimal representable positive term. Using fixnums for the latter, or flonums for the former limits the number of useful, positive radical terms. This implies that accuracy below \texttt{1e-10} (for \texttt{dyadic-cos1}) or below \texttt{1e-5} (for \texttt{dyadic-cos2}) is unattainable. Like in tail-recursion, which ensures that no memory is lost during recursion, we would like to be able to iterate continued radical as much as we need, with no loss of the accuracy. Of course, the final accuracy will always be limited by the precision of the output flonum (and indeed by the accuracy of the \texttt{sqrt} implementation). Our experiments with the trigonometric Taylor series and the \mac{CORDIC} indicate that these best-in-class algorithms do reach expected accuracy for \textsc{IEEE} \emph{double-precision} flonums (also called \texttt{binary64}) of about $2^{-52}/2\approx$\texttt{1e-16}, so we need to do better.

\begin{figure}[htb]
\label{Depth-limited apply}
\fbox{\hbox to .487\columnwidth{\vbox{\vspace{-.5em}{\small{\begin{flushleft}
{
{\let\oldtbf=\textbf\def\textbf#1{{\rm\oldtbf{#1}}}\noindent \texttt{\begin{tabular*}{0.90\linewidth}{l@{\extracolsep{\fill}}}
(\textbf{Define}\ (apply3\ fix:\ sig d)\\
\ (\textbf{Let}\ loop\ (fix:{\char91}s\ sig{\char93}\ {\char91}i\ 0{\char93}\ {\char91}a\ \textmd{\textit{:flo}}\ 0{\char93})\\
\ \ \ (\textbf{if}\ (>=\ i\ d)\ \\
\ \ \ \ \ (\textbf{+}\ 2\ a)\\
\ \ \ \ \ (loop\ (\textmd{\textit{rsh}}\ s\ 1)\ (+ i 1)\\
\ \ \ \ \ \ \ \ (\textbf{*}\ (>->\ s\ (\textit{and}\ \_\ 1)\ (\textmd{\textit{lsh}}\ \_\ 1)\\
\ \ \ \ \ \ \ \ \ \ \ \ \ \ \ \ \ \ \ \ \ \ \ \ \ \ \ \ \ (\textbf{-}\ 1\ \_))\\
\ \ \ \ \ \ \ \ \ \ \ (\textbf{sqrt}\ (\textbf{+}\ 2\ a)))\\
\ \ \ \ \ \ \ \ ))))\\
\end{tabular*}
}}
}\end{flushleft}
}}\vspace{-.5em}}}}
\label{Improved apply}\hspace{-.4em}
\fbox{\hbox to .487\columnwidth{\vbox{\vspace{-.5em}{\small{\begin{flushleft}
{
{\let\oldtbf=\textbf\def\textbf#1{{\rm\oldtbf{#1}}}\noindent \texttt{\begin{tabular*}{0.90\linewidth}{l@{\extracolsep{\fill}}}
(\textbf{Define}\ (apply4\ fix:\ sig')\\
\ (\textbf{Let}\ loop\ (fix:{\char91}s\ sig'{\char93}\ {\char91}a\ \textmd{\textit{:flo}}\ 0{\char93})\\
\ \ \ (\textbf{if}\ (<=\ s\ 3)\ \\
\ \ \ \ \ (\textbf{+}\ 2\ a)\\
\ \ \ \ \ (loop\ (\textmd{\textit{rsh}}\ s\ 1)\\
\ \ \ \ \ \ \ \ (\textbf{*}\ (>->\ s\ (\textit{and}\ \_\ 1)\ (\textmd{\textit{lsh}}\ \_\ 1)\\
\ \ \ \ \ \ \ \ \ \ \ \ \ \ \ \ \ \ \ \ \ \ \ \ \ \ \ \ \ (\textbf{-}\ 1\ \_))\\
\ \ \ \ \ \ \ \ \ \ \ (\textbf{sqrt}\ (\textbf{+}\ 2\ a)))\\
\ \ \ \ \ \ \ \ ))))\\
\end{tabular*}
}}
}\end{flushleft}
}}\vspace{-.5em}}}}
\vspace{-1em}\caption{{\color{black}\label{apply4}
}{\color{black}}Improved depth-limited \texttt{apply3} and final \texttt{apply4}}
\end{figure}

Going back to our very first equation (Pythagorean identity), however, we see that $\cos{\pi\over 4}={1\over 2}\sqrt{2}$. We observe that halving the radical can simplify (possibly vanishing) term(s) that are inside the radical. To see how this works out let us rewrite the whole expression as $s^k_0\sqrt{f_1+s^k_1\sqrt{f_2+s^k_2\sqrt{...}}}$ where $f_0=2^{-0}=1, f_1=2^{-1}, f_2=2^{-3}, f_i={{f_{i-1}^2}\over 2}=2^{-2^i+1} ({\rm for\ }i\ge 0)$. Let us also drop the omnipresent argument to $s^k_i(n)$ for simplicity of the formulas. Transforming our last equation from the beginning of section \ref{Development}, let us therefore halve the whole radical, bringing inside the radical a factor of $2^2=4$ to compensate:

$${s^k_0\over 2}\sqrt{{2^2\over 2}+2^2s^k_1\sqrt{{1\over 2*2^2}+s^k_2\sqrt{{1\over 2*(2*2^2)^2}+...}}}={s^k_0\over 2}\sqrt{2+s^k_1\sqrt{{4^2\over 2*2^2}+4^2 s^k_2\sqrt{{1\over 2*(2*2^2)^2}+...}}}$$

We see that in order to bring the re-balancing factor inside each nested radical we need to square the last factor, starting from 2.

$${s^k_0\over 2}\sqrt{2+s^k_1\sqrt{2+s^k_2\sqrt{{4^4\over 2*(2*4)^2}+4^4 s^k_3\sqrt{...}}}}={s^k_0\over 2}\sqrt{2+s^k_1\sqrt{2+s^k_2\sqrt{2+s^k_3\sqrt{2+...}}}}$$

Thus the iterated re-balancing of the factors $f_i,\forall i>0$ through multiplication by $g_0=2, g_1=2^2, g_i=2^{2^i} ({\rm for\ }i\ge 0)$ giving $g_i*f_i=2^{2^i-2^i+1}=2$ successfully brings the halving that causes terms to become smaller and smaller completely outside the radical. This leaves us with much simpler \texttt{apply4} that does not even need to perform squaring and uses a single constant factor $f_i=2, \forall i>0$, see Figure \ref{apply4}.

Note that both the depth of binary search as well as the signature are now represented by \texttt{sig'} where the initial value is 2 (\texttt{{\char35}b10} in binary). This initial value is our termination condition when we unwind the bit-mask by right-shifting. In the code above, however, we check for $s\le 3$ (the reason for this shall become apparent very soon). Before, the depth needed to be represented separately (by \texttt{lf}) since the mapping between $s^k_i(n)$ and \texttt{sig} is not \emph{bijective}. In fact, \texttt{sig} has a value of 0 for all ${n\over m}=2^{-k}, k\ge 1$. Although this is sufficient for identifying the signs given a value (i.e., the mapping is \emph{surjective}), we really need \emph{injectivity} in order to keep track of the tree depth and stop the tail recursion appropriately.

\begin{figure}[htb]
\label{Improved and generalised loop for dyadic rationals}
\fbox{\hbox to .99\columnwidth{\vbox{\vspace{-.5em}{\small{\begin{flushleft}
{
{\let\oldtbf=\textbf\def\textbf#1{{\rm\oldtbf{#1}}}\noindent \texttt{\begin{tabular*}{0.90\linewidth}{l@{\extracolsep{\fill}}}
(\textbf{Define}\ (dyadic-loop\ x\ \textmd{\textit{:rat}}\ sig'\ \textmd{\textit{:fix}})\\
(\textbf{cond}\ ({\char91}\textbf{=}\ x\ 0{\char93}\ (\textbf{*}\ 4\ (\textbf{-}\ 1\ (\textit{and}\ sig'\ 1))))\ \ \ \ \ \ \ \ \ \ \ \ \ \ \ \ \ \textit{;; exact }\\
\ \ \ \ \ \ ({\char91}\textbf{=}\ x\ 1{\char93}\ (\textbf{*}\ 4\ (\textit{and}\ sig'\ 1)))\ \ \ \ \ \ \ \ \ \ \ \ \ \ \ \ \ \ \ \ \ \ \ \textit{;; results}\\
(\textbf{else}\ (\textbf{Let}\ loop\ (fix:{\char91}n\ (\textbf{numerator}\ x){\char93}\\
\ \ \ \ \ \ \ \ \ \ \ \ \ \ \ \ \ \ \ \ {\char91}m\ (\textbf{denominator}\ x){\char93}\\
\ \ \ \ \ \ \ \ \ \ \ \ \ \ \ \ \ \ \ \ {\char91}s\ sig'{\char93})\\
\ (\textbf{Let*}\ ({\char91}m\ \textmd{\textit{:fix}}\ (\textmd{\textit{rsh}}\ m\ 1){\char93})\\
\ \ (\textbf{cond}\ ({\char91}\textbf{=}\ n\ m{\char93}\ (apply4\ s))\\
\ \ \ \ \ \ \ \ ({\char91}<\ n\ m{\char93}\ (loop\ n\ m\\
\ \ \ \ \ \ \ \ \ \ \ \ \ \ \ \ \ \ \ \ (>->\ s\ (\textmd{\textit{lsh}}\ \_\ 1)\ \ \ \\
\ \ \ \ \ \ \ \ \ \ \ \ \ \ \ \ \ \ \ \ \ \ \ (\textmd{\textit{bit-count}}\ s)\ \ \ \\
\ \ \ \ \ \ \ \ \ \ \ \ \ \ \ \ \ \ \ \ \ \ \ (\textit{and}\ \_\ 1)\ \ \ \ \ \ \ \ \ \ \ \ \ \ \ \ \ \ \ \ \ \ \ \textit{;; bit-counting}\\
\ \ \ \ \ \ \ \ \ \ \ \ \ \ \ \ \ \ \ \ \ \ \ (\textbf{-}\ 1\ \_)\ \ \ \ \ \ \ \ \ \ \ \ \ \ \ \ \ \ \ \ \ \ \ \ \ \ \textit{;; logic reversed}\\
\ \ \ \ \ \ \ \ \ \ \ \ \ \ \ \ \ \ \ \ \ \ \ (\textit{or}\ \_\ {\char91}\textbf{Nth}\ 4{\char93})\ \ \\
\ \ \ \ \ \ \ \ \ \ \ \ \ \ \ \ \ \ \ \ )))\\
\ \ \ \ \ \ \ \ ({\char91}>\ n\ m{\char93}\ (loop\ (\textbf{-}\ n\ m)\ m\\
\ \ \ \ \ \ \ \ \ \ \ \ \ \ \ \ \ \ \ \ (>->\ s\ (\textmd{\textit{lsh}}\ \_\ 1)\\
\ \ \ \ \ \ \ \ \ \ \ \ \ \ \ \ \ \ \ \ \ \ \ (\textmd{\textit{bit-count}}\ s)\\
\ \ \ \ \ \ \ \ \ \ \ \ \ \ \ \ \ \ \ \ \ \ \ (\textit{and}\ \_\ 1)\ \ \ \ \ \ \ \ \ \ \ \ \ \ \ \ \ \ \ \ \ \ \ \textit{;; bit-counting}\\
\ \ \ \ \ \ \ \ \ \ \ \ \ \ \ \ \ \ \ \ \ \ \ (\textit{or}\ \_\ {\char91}\textbf{Nth}\ 3{\char93})\ \ \ \ \ \ \ \ \ \ \ \ \ \ \ \ \ \ \textit{;; logic reversed}\\
\ \ \ \ \ \ \ \ \ \ \ \ \ \ \ \ \ \ \ \ )))))))))\\
\end{tabular*}
}}
}\end{flushleft}
}}\vspace{-.5em}}}}
\vspace{-2em}\caption{{\color{black}\label{loop}
}{\color{black}}Improved and generalised loop for dyadic rationals}
\end{figure}

Let us use \texttt{sig'} in Figure \ref{loop} then, which is a primed version of \texttt{sig} we discussed above. It implies that the bit-counting logic for the main loop needs to be reversed (since there is now an extra bit in the most significant position). Next to this, we can now simplify calculation of $s^k_i(n)$ even more by making use of $1-2(s\bmod 2)$ to represent the mapping $\{1\mapsto -1,0\mapsto 1\}$. Similarly in \texttt{dyadic-loop} function, we can rewrite the ``update'' of \texttt{sig} for the next iteration accordingly, making use of the representation $\{1\mapsto 0,0\mapsto 1\}\equiv 1-s\bmod 2$:

{\small{\begin{flushleft}
{
{\let\oldtbf=\textbf\def\textbf#1{{\rm\oldtbf{#1}}}\noindent \texttt{\begin{tabular*}{0.90\linewidth}{l@{\extracolsep{\fill}}}
(>->\ s\ (\textmd{\textit{lsh}}\ \_\ 1)\\
\ \ \ \ (\textit{or}\ \_\ (\textbf{if}\ (\textmd{\textit{even?}}\ (\textmd{\textit{bit-count}}\ s))\\
\ \ \ \ \ \ \ \ \ \ \ \ \ \ 1\ 0)))\\
\end{tabular*}
}}
}\end{flushleft}
}}
\noindent $-\{$strength-reduce$\}\rightarrow$
{\small{\begin{flushleft}
{
{\let\oldtbf=\textbf\def\textbf#1{{\rm\oldtbf{#1}}}\noindent \texttt{\begin{tabular*}{0.90\linewidth}{l@{\extracolsep{\fill}}}
(>->\ s\ \ \ \ \ \ \ \ \ \ \ \ \ \ \textit{;; {\char91}Nth 5{\char93}}\\
\ \ \ \ (\textmd{\textit{lsh}}\ \_\ 1)\ \ \ \ \ \ \ \textit{;; {\char91}Nth 4{\char93}}\\
\ \ \ \ (\textmd{\textit{bit-count}}\ s)\ \ \ \textit{;; {\char91}Nth 3{\char93}}\\
\ \ \ \ (\textit{and}\ \_\ 1)\ \ \ \ \ \ \ \textit{;; {\char91}Nth 2{\char93}}\\
\ \ \ \ (\textbf{-}\ 1\ \_)\ \ \ \ \ \ \ \ \ \ \textit{;; {\char91}Nth 1{\char93}}\\
\ \ \ \ (\textit{or}\ \_\ {\char91}\textbf{Nth}\ 4{\char93}))\ \textit{;; refer to the above sub-expressions 1 and 4}\\
\end{tabular*}
}}
}\end{flushleft}
}}

Combining all these improvements together we arrive at our final parsimonious iterative, mostly control-flow free driver for our target function, which now inputs a \texttt{rat:} annotated dyadic rational \texttt{x} and the initial mask \texttt{sig'}. Even though rational arithmetic is specified in \mac{R5RS}, not all implementations include them by default. With macros, however, it is not difficult to extend Bigloo to handle rational numbers seamlessly (see section \ref{Tangential: rational arithmetic} where we show how \texttt{sqrt} can be overloaded for rational inputs).

A keen reader will have noticed that \texttt{apply4} now stops earlier and does not apply the final \texttt{sqrt} (and halving). For the same reason, in the \texttt{dyadic-loop}, the exact result is multiplied by 4. The answer as to why was to be expected: we want to be able to calculate functions other than just $\cos{n\pi\over 2m}$ of course!

\subsubsection{A surprising generalisation: Sine
}
\label{A surprising generalisation: Sine}

Keeping track of the signs in the sequence $s^k_i, i\in[0,1,2 ...]$ and interpreting bit-mask \texttt{sig} as such a sequence (using 1 to represent $-1$ and 0 to represent $1$) has another benevolent side effect: we can actually represent the $\sin{}$ function by exact same algorithm as for $\cos{}$, just using initial value of $s^k_0=-1$.

This corresponds to the initial value of \texttt{sig'={\char35}b11=3} instead of \texttt{{\char35}b10=2}, incidentally explaining why we used $\le 3$ to check for termination in \texttt{apply4} from Figure \ref{apply4}. This is because in general, $\sin{x}=\sqrt{1-\cos^2{x}}=\sqrt{1-{1\over 2^2}(2\pm\sqrt{...})}={1\over 2}\sqrt{4-{4\over 2^2}(2\pm\sqrt{...})}={1\over 2}\sqrt{2\mp\sqrt{...}}$, so simply inverting the sign under the first radical will do the trick.
Equipped with the higher-order machinery from section \ref{Tangential: composition transformers} which introduces a syntactic composition operator ($\circ$) and a currying form ($\kappa$), we can now attack other trigonometric functions by playing with initial values of \texttt{sig'} and values returned from \texttt{dyadic-loop/apply4}. For example, squared results are simply obtained by \emph{not} performing final \texttt{sqrt}. Expressions for $\tan{}$ and $\cot{}$ are well-known from elementary trigonometry.

{\small{\begin{flushleft}
{
{\let\oldtbf=\textbf\def\textbf#1{{\rm\oldtbf{#1}}}\noindent \texttt{\begin{tabular*}{0.90\linewidth}{l@{\extracolsep{\fill}}}
(\textbf{Define}\ (dyadic-cos\ x)\ ((∘\ \textbf{half}\ \textbf{sqrt})\ (dyadic-loop\ x\ {\char35}b10)))\\
(\textbf{Define}\ (dyadic-sin\ x)\ ((∘\ \textbf{half}\ \textbf{sqrt})\ (dyadic-loop\ x\ {\char35}b11)))\\
(\textbf{Define}\ (dyadic-tan\ x)\ ((∘\ \textbf{sqrt}\ \textbf{sub1}\ \textbf{reciprocal}\ {\char91}κ\ \textbf{/}\ \_\ 4{\char93})\ (dyadic-loop\ x\ {\char35}b10)))\\
(\textbf{Define}\ (dyadic-cot\ x)\ ((∘\ \textbf{sqrt}\ \textbf{sub1}\ \textbf{reciprocal}\ {\char91}κ\ \textbf{/}\ \_\ 4{\char93})\ (dyadic-loop\ x\ {\char35}b11)))\\
(\textbf{Define}\ (dyadic-cos²\ x)\ ({\char91}κ\ \textbf{/}\ \_\ 4{\char93}\ (dyadic-loop\ x\ {\char35}b10)))\\
(\textbf{Define}\ (dyadic-sin²\ x)\ ({\char91}κ\ \textbf{/}\ \_\ 4{\char93}\ (dyadic-loop\ x\ {\char35}b11)))\\
(\textbf{Define}\ (dyadic-tan²\ x)\ ((∘\ \textbf{sub1}\ \textbf{reciprocal}\ {\char91}κ\ \textbf{/}\ \_\ 4{\char93})\ (dyadic-loop\ x\ {\char35}b10)))\\
(\textbf{Define}\ (dyadic-cot²\ x)\ ((∘\ \textbf{sub1}\ \textbf{reciprocal}\ {\char91}κ\ \textbf{/}\ \_\ 4{\char93})\ (dyadic-loop\ x\ {\char35}b11)))\\
\end{tabular*}
}}
}\end{flushleft}
}}

\subsection{Inverse trigonometric functions
}
\label{Inverse trigonometric functions}

\begin{figure}[htb]
\label{First attempt at inverse dyadic cosine}
\fbox{\hbox to .99\columnwidth{\vbox{\vspace{-.5em}{\small{\begin{flushleft}
{
{\let\oldtbf=\textbf\def\textbf#1{{\rm\oldtbf{#1}}}\noindent \texttt{\begin{tabular*}{0.90\linewidth}{l@{\extracolsep{\fill}}}
(\textbf{Define}\ (dyadic-cos-1\ x\ \textmd{\textit{:flo}}\ ss\ \textmd{\textit{:fix}}\ *eps*\ \textmd{\textit{:flo}})\\
(\textbf{cond}\ ({\char91}\textbf{=}\ x\ 0.{\char93}\ (\textbf{-}\ 1\ ss))\ \ \ \ \ \ \ \ \ \ \ \ \ \ \ \ \ \ \ \ \ \ \ \ \ \ \ \ \textit{;; exact}\\
\ \ \ \ \ \ ({\char91}\textbf{=}\ x\ 1.{\char93}\ ss)\ \ \ \ \ \ \ \ \ \ \ \ \ \ \ \ \ \ \ \ \ \ \ \ \ \ \ \ \ \ \ \ \ \ \textit{;; results}\\
\ \ \ \ \ \ (\textbf{else}\\
(\textbf{Let}\ loop\ ({\char91}y\ \textmd{\textit{:rat}}\ {\char35}r:1/2{\char93}\\
\ \ \ \ \ \ \ \ \ \ \ {\char91}s\ :fix\ ss{\char93}\\
\ \ \ \ \ \ \ \ \ \ \ {\char91}d\ :fix\ 0{\char93})\\
\ (\textbf{Let}\ ({\char91}m\ \textmd{\textit{:fix}}\ (>->\ (\textbf{denominator}\ y)\ \textmd{\textit{:fix}}\ (\textmd{\textit{lsh}}\ \_\ 1)){\char93}\\
\ \ \ \ \ \ \ {\char91}xe\ \textmd{\textit{:flo}}\ ((∘\ \textbf{half}\ \textbf{sqrt})\ (apply3\ s\ d)){\char93}\\
\ \ \ \ \ \ \ {\char91}sss\ (>->\ ss\ (\textmd{\textit{lsh}}\ \_\ 1)\ (\textbf{-}\ 1\ \_)\ (toflo\ \_)){\char93})\\
\ \ (\textbf{cond}\ ({\char91}<\ (\textbf{abs}\ (\textbf{-}\ x\ xe))\ *eps*{\char93}\ y)\\
\ \ \ \ \ \ \ \ ({\char91}<\ (\textbf{*}\ sss\ x)\ (\textbf{*}\ sss\ xe){\char93}\\
\ \ \ \ \ \ \ \ \ (loop\ (\textbf{+}\ y\ {\char35}r:1/m)\\
\ \ \ \ \ \ \ \ \ \ \ \ (>->\ s\ (\textmd{\textit{lsh}}\ \_\ 1)\\
\ \ \ \ \ \ \ \ \ \ \ \ \ \ \ (\textmd{\textit{bit-count}}\ s)\\
\ \ \ \ \ \ \ \ \ \ \ \ \ \ \ (\textit{and}\ \_\ 1)\\
\ \ \ \ \ \ \ \ \ \ \ \ \ \ \ (\textbf{-}\ 1\ \_)\\
\ \ \ \ \ \ \ \ \ \ \ \ \ \ \ (\textit{or}\ \_\ {\char91}\textbf{Nth}\ 4{\char93}))\\
\ \ \ \ \ \ \ \ \ \ \ \ (\textbf{+}\ d\ 1)\\
\ \ \ \ \ \ \ \ \ \ \ \ ))\\
\ \ \ \ \ \ \ \ ({\char91}>\ (\textbf{*}\ sss\ x)\ (\textbf{*}\ sss\ xe){\char93}\ \\
\ \ \ \ \ \ \ \ \ (loop\ (\textbf{-}\ y\ {\char35}r:1/m)\\
\ \ \ \ \ \ \ \ \ \ \ \ (>->\ s\ (\textmd{\textit{lsh}}\ \_\ 1)\\
\ \ \ \ \ \ \ \ \ \ \ \ \ \ \ (\textmd{\textit{bit-count}}\ s)\\
\ \ \ \ \ \ \ \ \ \ \ \ \ \ \ (\textit{and}\ \_\ 1)\\
\ \ \ \ \ \ \ \ \ \ \ \ \ \ \ (\textit{or}\ \_\ {\char91}\textbf{Nth}\ 3{\char93}))\\
\ \ \ \ \ \ \ \ \ \ \ \ (\textbf{+}\ d\ 1)\\
\ \ \ \ \ \ \ \ \ \ \ \ ))))))))\\
\end{tabular*}
}}
}\end{flushleft}
}}\vspace{-.5em}}}}
\vspace{-2em}\caption{{\color{black}\label{cos-1}
}{\color{black}}First attempt at inverse dyadic cosine}
\end{figure}

There are many approaches for implementing a reverse-mode algorithm provided an implementation of a forward-mode bijective function. Most straightforward is to search for the value in the domain provided a value from the range. Since both $\sin{x}$ and $\cos{x}$ are monotonic in the interval $x\in[0,{\pi\over 2}]$, we could simply apply a search procedure like the one depicted in Figure \ref{cos-1}. On first sight, it seems to be efficient, using binary search that dichotomises the input domain $[0,1]$ according to the values returned by the forward function given the corresponding search key (i.e, our dyadic rational \texttt{{\char35}r:n/m} representing the argument), which gets successively approximated according to the deviation from the target value returned by the forward function evaluation.

However, a proper search procedure should not insist on extra $O(\log{1\over p})$ evaluations of a continued radical (where $p={1\over m}=2^{-k}$ is the precision of the input value). Even though with each new level we can stop evaluation at $p$ (corresponding to the depth of $\log{1\over p}=k$), the total cost still amounts to $1+2+3+...+k=O(k^2)$ which as we know grows very quickly as $p\rightarrow 0$. Besides, we are seeking a general method for all trigonometric inverses, not just one for $\cos^{-1}{x}$ (a value returned by \texttt{apply3} needs to be translated to actual forward function target value, using \texttt{($\circ$ half sqrt)}). We would really like to fuse the binary search together with the evaluation of each corresponding target value used to drive the decision process, and get rid of the overhead that depends on the argument.

\begin{figure}[htb]
\label{Fused and generalised inverse on dyadic rationals}
\fbox{\hbox to .99\columnwidth{\vbox{\vspace{-.5em}{\small{\begin{flushleft}
{
{\let\oldtbf=\textbf\def\textbf#1{{\rm\oldtbf{#1}}}\noindent \texttt{\begin{tabular*}{0.90\linewidth}{l@{\extracolsep{\fill}}}
(\textbf{Define}\ (dyadic-find5\ x\ \textmd{\textit{:flo}}\ ss\ \textmd{\textit{:fix}}\ *eps*\ \textmd{\textit{:flo}})\\
(\textbf{cond}\ ({\char91}\textbf{=}\ x\ 0.{\char93}\ (\textbf{-}\ 1\ ss))\ \ \ \ \ \ \ \ \ \ \ \ \ \ \ \ \ \ \ \ \ \ \ \ \ \ \textit{;; exact}\\
\ \ \ \ \ \ ({\char91}\textbf{=}\ x\ 4.{\char93}\ ss)\ \ \ \ \ \ \ \ \ \ \ \ \ \ \ \ \ \ \ \ \ \ \ \ \ \ \ \ \ \ \ \ \textit{;; results}\\
(\textbf{else}\ (\textbf{Let}\ loop\ ({\char91}y\ \textmd{\textit{:rat}}\ {\char35}r:1/2{\char93}\\
\ \ \ \ \ \ \ \ \ \ \ \ \ \ \ \ {\char91}x\ \textmd{\textit{:flo}}\ x{\char93}\\
\ \ \ \ \ \ \ \ \ \ \ \ \ \ \ \ {\char91}s\ :fix\ ss{\char93})\\
\ (\textbf{Let}\ ({\char91}m\ \textmd{\textit{:fix}}\ (>->\ (\textbf{denominator}\ y)\ \textmd{\textit{:fix}}\ (\textmd{\textit{lsh}}\ \_\ 1)){\char93})\\
\ \ (\textbf{cond}\ ({\char91}<\ (\textbf{abs}\ (\textbf{-}\ x\ 2))\ *eps*{\char93}\ y)\\
\ \ \ \ \ \ \ \ ({\char91}<\ x\ 2.{\char93}\\
\ \ \ \ \ \ \ \ \ (loop\ (\textbf{+}\ y\ (\textbf{*}\ (>->\ s\ (\textmd{\textit{bit-count}}\ \_)\\
\ \ \ \ \ \ \ \ \ \ \ \ \ \ \ \ \ \ \ \ \ \ \ \ \ \ (\textit{and}\ \_\ 1)\\
\ \ \ \ \ \ \ \ \ \ \ \ \ \ \ \ \ \ \ \ \ \ \ \ \ \ (\textmd{\textit{lsh}}\ \_\ 1)\\
\ \ \ \ \ \ \ \ \ \ \ \ \ \ \ \ \ \ \ \ \ \ \ \ \ \ (\textbf{-}\ 1\ \_))\\
\ \ \ \ \ \ \ \ \ \ \ \ \ \ \ \ \ \ \ \ \ \ \ {\char35}r:1/m))\\
\ \ \ \ \ \ \ \ \ \ \ \ (>->\ x\ (\textbf{-}\ \_\ 2)\ (\textbf{sqr}\ \_))\\
\ \ \ \ \ \ \ \ \ \ \ \ (>->\ s\ (\textmd{\textit{lsh}}\ \_\ 1)\ (\textit{or}\ \_\ 1))\\
\ \ \ \ \ \ \ \ \ \ \ \ ))\\
\ \ \ \ \ \ \ \ ({\char91}>\ x\ 2.{\char93}\ \\
\ \ \ \ \ \ \ \ \ (loop\ (\textbf{-}\ y\ (\textbf{*}\ (>->\ s\ (\textmd{\textit{bit-count}}\ \_)\\
\ \ \ \ \ \ \ \ \ \ \ \ \ \ \ \ \ \ \ \ \ \ \ \ \ \ (\textit{and}\ \_\ 1)\\
\ \ \ \ \ \ \ \ \ \ \ \ \ \ \ \ \ \ \ \ \ \ \ \ \ \ (\textmd{\textit{lsh}}\ \_\ 1)\\
\ \ \ \ \ \ \ \ \ \ \ \ \ \ \ \ \ \ \ \ \ \ \ \ \ \ (\textbf{-}\ 1\ \_))\\
\ \ \ \ \ \ \ \ \ \ \ \ \ \ \ \ \ \ \ \ \ \ \ {\char35}r:1/m))\\
\ \ \ \ \ \ \ \ \ \ \ \ (>->\ x\ (\textbf{-}\ \_\ 2)\ (\textbf{sqr}\ \_))\\
\ \ \ \ \ \ \ \ \ \ \ \ (>->\ s\ (\textmd{\textit{lsh}}\ \_\ 1)\ (\textit{or}\ \_\ 0))\\
\ \ \ \ \ \ \ \ \ \ \ \ ))))))))\\
\end{tabular*}
}}
}\end{flushleft}
}}\vspace{-.5em}}}}
\vspace{-2em}\caption{{\color{black}\label{find5}
}{\color{black}}Fused and generalised inverse on dyadic rationals}
\end{figure}

Looking again at our radical from section \ref{Control and representation optimisations}, we see that repeated squaring can undo the steps performed by the forward-mode trigonometric algorithm. Since the argument is uniquely determined by the signature $s^k_i$ (refer to Figure \ref{dy16}), this sequence of signs is exactly what we seek when computing an inverse. It is natural to accomplish this by performing the steps of the forward algorithm in reverse:

\begin{enumerate}

\item initial remainder is set to the input value, doubled and squared: $(\circ\rm\ half\ sqrt)^{-1}=(\circ\rm\ sqr\ double)$

\item at every un-nesting, determine:
\begin{enumerate}

\item if the remainder is approximately equal to 2, terminate the search

\item if less than 2, the sign $s^k_i=-1$

\item if greater than 2, the sign is $s^k_i=1$

\item In either case, 2 is subtracted and the remainder is squared

\end{enumerate}

\end{enumerate}

The argument is adjusted according to the bit-counting logic, taking binary search history into account. We can do all of this precisely because $\sqrt{x}\ge 0,\forall x\ge 0$ always, allowing the conditional to uniquely determine the sign sequence (see Figure \ref{find5}).
And indeed, a repeated application of \texttt{sqr} instead of repeated application of \texttt{sqrt}, is essentially ``unwinding'' the nested radical and therefore correctly computing the ${n\over m}={2\cos^{-1}{x}\over\pi}$ with provably no superfluous computational steps performed.

{\small{\begin{flushleft}
{
{\let\oldtbf=\textbf\def\textbf#1{{\rm\oldtbf{#1}}}\noindent \texttt{\begin{tabular*}{0.90\linewidth}{l@{\extracolsep{\fill}}}
(\textbf{def-syntax}\ *eps*\ 1e-10)\\
(\textbf{Define}\ {\char91}dyadic-acos\ x{\char93}\ (dyadic-find5\ ((∘\ \textbf{sqr}\ \textbf{double})\ x)\ 0\ *eps*))\\
(\textbf{Define}\ {\char91}dyadic-asin\ x{\char93}\ (dyadic-find5\ ((∘\ \textbf{sqr}\ \textbf{double})\ x)\ 1\ *eps*))\\
(\textbf{Define}\ {\char91}dyadic-atan\ x{\char93}\ (dyadic-find5\ ((∘\ \textbf{sqr}\ \textbf{double}\ \textbf{sqrt}\ \textbf{reciprocal}\ \textbf{add1}\ \textbf{sqr})\ x)\ 0\ *eps*))\\
(\textbf{Define}\ {\char91}dyadic-acot\ x{\char93}\ (dyadic-find5\ ((∘\ \textbf{sqr}\ \textbf{double}\ \textbf{sqrt}\ \textbf{reciprocal}\ \textbf{add1}\ \textbf{sqr})\ x)\ 1\ *eps*))\\
\end{tabular*}
}}
}\end{flushleft}
}}

Concluding, we notice in passing that \texttt{($\circ$ sqr double sqrt reciprocal add1 sqr)$\equiv$($\circ$ {\char91}$\kappa$ / 4 \_{\char93} add1 sqr)} so our final inverse functions can all be rewritten to avoid using the \texttt{sqrt}. This completes the exposition of the complementary relationship between trigonometric functions that we expect from the mathematical point of view: forward-mode functions use the \texttt{sqrt} while the reverse-mode functions use the \texttt{sqr}. It would be not surprising if the implementation of either side of this dualism would be derived mechanically from the other side, automatically proving correctness of the approach.

\section{Evaluation}
\label{Evaluation}

Because trigonometric functions are symmetric, a simple approach to extend the domain for our forward-mode to $[0,2\pi]$ is to record the \emph{quadrature} of the argument as a pair of signs (1st quadrant encoded as +/+, 2nd quadrant as -/+, 3rd quadrant as -/- and 4th as +/-). The first (\texttt{inv}) identifies the sign of $\cos{x}$ as $x$ circles the quadrants in sequence. The second sign (\texttt{mir}) acts similarly for the $\sin{}$ function.

{\small{\begin{flushleft}
{
{\let\oldtbf=\textbf\def\textbf#1{{\rm\oldtbf{#1}}}\noindent \texttt{\begin{tabular*}{0.90\linewidth}{l@{\extracolsep{\fill}}}
(\textbf{Define}\ (quadrature\ x\ \textmd{\textit{:rat}}\ inv\ \textmd{\textit{:fix}}\ mir\ \textmd{\textit{:fix}})\\
\ \ \ \ \ (\textbf{cond}\ ((\textbf{or}\ {\char91}>\ x\ {\char35}r:1/4{\char93}\ {\char91}<\ x\ {\char35}r:-1/4{\char93})\ \\
\ \ \ \ \ \ \ \ \ \ \ \ (quadrature\ (\textbf{-}\ x\ (\textbf{*}\ (\textbf{sign}\ x)\ {\char35}r:1/2))\ (\textmd{\textit{neg}}\ inv)\ (\textmd{\textit{neg}}\ mir)))\\
\ \ \ \ \ \ \ \ \ \ \ ({\char91}<\ x\ 0{\char93}\ (quadrature\ (\textmd{\textit{neg}}\ x)\ inv\ (\textmd{\textit{neg}}\ mir)))\\
\ \ \ \ \ \ \ \ \ \ \ (\textbf{else}\ (\textmd{\textit{values}}\ x\ inv\ mir))))\\
(\textbf{Define}\ (Cos\ x\ \textmd{\textit{:rat}})\\
\ \ \ \ \ (\textbf{let-values}\ ({\char91}(y\ inv\ mir)\ (quadrature\ x\ 1\ 1){\char93})\\
\ \ \ \ \ \ \ (\textbf{*}\ inv\ (dyadic-cos\ (\textbf{*}\ y\ 4)))))\\
(\textbf{Define}\ (Sin\ x\ \textmd{\textit{:rat}})\\
\ \ \ \ \ (\textbf{let-values}\ ({\char91}(y\ inv\ mir)\ (quadrature\ x\ 1\ 1){\char93})\\
\ \ \ \ \ \ \ (\textbf{*}\ mir\ (dyadic-sin\ (\textbf{*}\ y\ 4)))))\\
\end{tabular*}
}}
}\end{flushleft}
}}

These functions use the \emph{multiple-value} facility, specified in \mac{SRFI} {\char35}11 \cite{SRFI11}, which allows the caller to simultaneously receive more than one return value from the callee. The latter constructs such a multiple-value using the \texttt{values} form.
Similarly, the signs of the input domain can be used to expand the range of the result for reverse-mode trigonometric functions to their appropriate values ($[{-\pi\over 2},{\pi\over 2}]$ for $\arcsin{}$ and $[0,\pi]$ for $\arccos{}$, $({-\pi\over 2},{\pi\over 2})$ for $\arctan{}$ and $(0,\pi)$ for $acot{}$).

In order to properly evaluate accuracy and performance of our algorithms and ensure fair comparison to baseline \CEE/ implementations, we have re-implemented them the improved version of the above algorithms using \CEE/, taking the above Scheme versions as a specification (and as a proof of correctness). The use of multi-cycle \texttt{sqrt} in forward-mode trigonometry prevents efficient leverage of the approach. For inverse functions, however, we require \emph{squaring}, which is much less complex than \texttt{sqrt} or a a full multiplier hardware-wise (see \cite{deshpande2010squaring}, \cite{Sethi2015}).
All tests used \mac{GCC} 7.2.0 running on \texttt{x86\_64-linux-gnu} architecture (Ubuntu 17.10 4-way Intel(R) Core(TM) i7-4600U CPU @ 2.10GHz). Used compiler options were: \texttt{-static -Ofast -mabm -ffast-math -fschedule-insns2}. We used \texttt{1e7} random \textsc{IEEE} \texttt{double} values as the basis for comparison.

\subsection{Accuracy
}
\label{Accuracy}

We mentioned above that \textsc{IEEE} double-precision flonums support accuracies of down to \texttt{1e-16}. Our implementation of basic trigonometric functions reaches this accuracy (going as low as \texttt{8e-17} for $\cos{}$ and $\sin{}$). Their inverses $\arccos{}$ and $\arcsin{}$ return exact (rational) results. To reach similar accuracy of \texttt{5e-17} (even on the interval $x\in[0,{\pi\over 2}]$), a Taylor series for $\cos{x}$ would need to be expanded to the 10th term (requiring calculation of $x^{20}$, as well as a division by $20!$). Unfortunately, even when applying the Horner's rule, a series of 10 square operations, accompanied by 10 full multipliers and significantly complicated dividers would be needed to match the accuracy of the presented dyadic cosine.

\begin{figure}[htb]
\label{RMS and MXM errors}
\noindent
\includegraphics[width=0.49\linewidth]{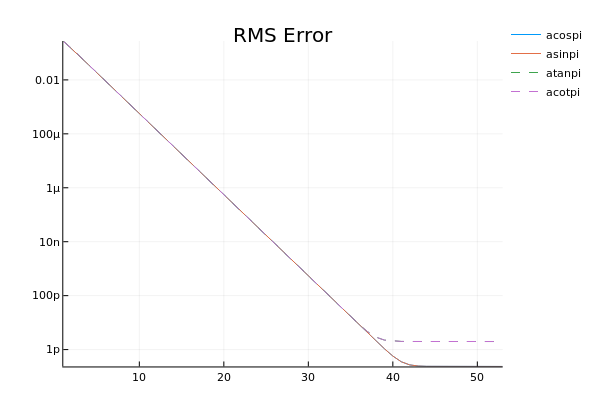}
\noindent
\includegraphics[width=0.49\linewidth]{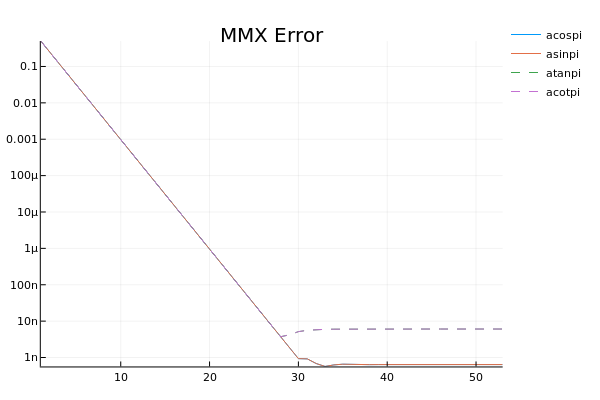}
\caption{{\color{black}\label{error}
}{\color{black}}RMS and MXM errors (for reverse-mode trigonometric functions)}\end{figure}

Evaluation at a dyadic rational with depth equal to its precision still gives accurate results, while for regular iterative algorithms the performance has characteristics that follow the law of diminishing returns.
One sees from Figure \ref{error} that $\arctan{}$ and $\rm arccot{}$ do reach the error floor at $\approx$ \texttt{4e-24} while  $\arccos{}$ and $\arcsin{}$ can yet be $\approx$ 2 orders of magnitude more accurate (as they don't use reciprocals).

Similarly, for the \mac{CORDIC}, at least 45 (for reverse-mode trigonometry) or 52 (for forward-mode trigonometry) iterations would be required to reach comparable accuracy of \texttt{1e-16}, since on each iteration, the \mac{CORDIC} produces a single bit of the result.
In our experiments we have observed that our algorithms for $\tan{}$ and $\cot{}$ turn out to be order-3 less accurate than the other two trigonometric functions. We attribute this to the use of floating-point \texttt{reciprocal} in their implementations.
Figure \ref{error} shows that \mac{RMS} error (defined as $\sqrt{\overline{(target-expect)^2}}$) exhibits linear scaling up to $\approx$ 40 iterations while \mac{MXM} error (defined as $\max{|target-expect|}$) exhibits linear scaling up to $\approx$ 30 iterations, which is often acceptable in embedded applications.

\subsection{Performance
}
\label{Performance}

Both forward and reverse trigonometric functions appear to require a logarithmic $O(\log{m})=O(k)$ number of iterations (thanks to the binary search). The use of only basic arithmetic operations, and exclusive use of tail-recursion implies that presented algorithms map well to embedded hardware and software implementations. Presented method of performing the \texttt{sqrt} and existing approaches for \texttt{sqr} together imply that the combined performance of our dyadic trigonometry is expected to be reasonably good.

Comparing run-times of this implementation when executing the full scale of $2^{20}$ dyadic rationals considered in this paper (compiled to native code via {\CEE/} using Bigloo 4.1a), we see that it is only $2.74$ times slower than standard implementations (originating in the C math library). For dyadics with precision $2^{-10}$ the slowdown is only $1.57$ (for $k=5$ it is just $1.2$, and so on with the asymptotic limit of $1.06$). We believe this overhead can be addressed by rewriting the algorithm in {\CEE/}, if needed (which is not a huge undertaking given the effort we spent to make it embeddable).
The accuracy and performance is achieved at the cost of generality: our trigonometric functions only accept dyadic rationals and (in their presented form) may completely fail for other rationals. E.g., for ${n\over m}={5\over 2^k-1}$ our \texttt{dyadic-loop} may even fail to terminate!

\begin{figure}[htb]
\label{Performance and speedup wrt. the baseline}
\noindent
\includegraphics[width=0.49\linewidth]{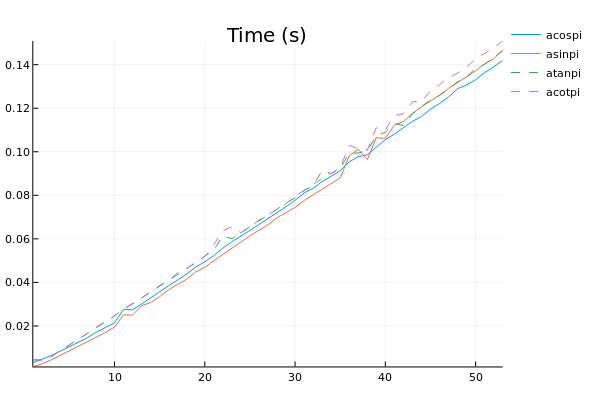}
\noindent
\includegraphics[width=0.49\linewidth]{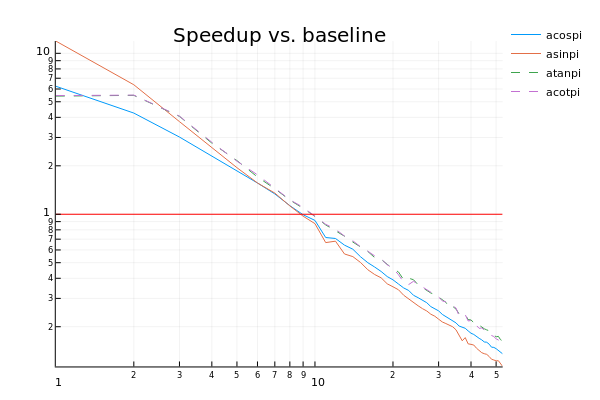}
\caption{{\color{black}\label{perf}
}{\color{black}}Performance and speedup wrt. the baseline (for reverse-mode trigonometric functions)}\end{figure}

From Figure \ref{perf} one can observe that unlike baseline implementations (which require constant time for any argument), our algorithm can be scaled, depending on required precision. With our current {\CEE/} implementation, we observe that at around 9 iterations, one can generate a result which is still accurate to $\approx$ 3 decimal digits faster than the off-the-shelf implementation.

\subsection{Related work
}
\label{Related work}

Trigonometry has been in widespread use since ancient times. It still forms the basis of many branches of science and engineering. Complex numbers, for example, naturally correspond to conjoined trigonometric functions. Many approaches to compute trigonometric functions have been tried. Tables \cite{boyer2011history} have been created in Babylon, during the classic period, in the Arabic world, in India and China. \mac{CORDIC} methods \cite{volder1956binary} rely on tabling of the values $x$ for which corresponding $\tan{x}=2^{-k},k\in\nats$. Using Taylor/Maclaurin series is also popular, however, applying these beyond just a few terms can quickly become costly. Methods that combine tabling, series expansions or polynomial interpolation are also known \cite{gal1986computing}.

Efficient methods to compute \texttt{sqrt} and derived functions have been extensively studied too \cite{abramowitz1964handbook}. Computer arithmetic in general and \textsc{IEEE} \texttt{binary64} format in particular seem to be suitable for efficient implementations of \texttt{sqrt} \cite{kushner2002wizardry}.
Although the basic idea to express basic trigonometric functions using radicals is not new and has appeared for example in \cite{Wolfram} and in \cite{servi2003nested}, the exact algorithm presented in this paper was not generally known (at least, to the knowledge available to us). Even more so, our approach to fuse binary search together with the evaluation of inverses of basic trigonometric functions on dyadic rationals is new, as far as we can judge.

\section{Conclusions}
\label{Conclusions}

We believe our approach is radically different in that it brings together two seemingly unrelated functions (\texttt{sqrt} and \texttt{cos}) and expresses the latter in terms of the former. This application of functional programming is beneficial in showing that much good can materialise when one thinks algorithmically, in addition to thinking mathematically. In particular, one can derive an algorithm for the inverse quite straightforwardly given the algebraic representation of the forward calculation.

Comparing Taylor/Maclaurin series of forward and reverse trigonometric functions one sees very few similarities. We believe that our approach to derive the algorithm (and prove correct) the implementation of trigonometric functions and their inverses deserves some attention.

Threading macros provide powerful syntax to automatically generate complex applicative forms, where bindings are propagated to an expression in a deeply nested program structure. This is more general than traditional (point-free) function composition in that it does not require concatenative style programming ala Forth, with \texttt{dup}, \texttt{curry}, \texttt{uncurry}, sorting of the argument lists into the appropriate order etc., for each of the sub-expressions.

Our algorithms exhibit decent performance mainly because of the use of type-specialised operators and library functions. Dynamic typing and dispatch may be fine to start with, but in order to match the performance of {\CEE/}, any functional language targeting numerical and scientific computing needs to address both of these aspects.

We use the application of dyadic rational trigonometry to show the ease with which Scheme's weak hygiene may be subverted to suit somewhat more advanced (albeit esoteric to some) desires: sectioning, overloading, threading, extensions to the number tower. All these, together with composition transformers are beneficial as core components of an algorithmic \mac{DSL}. This improves the exposition of the algorithm as well as elucidates basic mathematical relationships between the algorithms for computation of trigonometric functions and their inverses. Although especially useful for this particular domain, tail-recursive style of Scheme in general is excellent in prototyping of embedded scientific and numerical algorithms.

\subsection{Future work
}
\label{Future work}

Continued radicals are similar in spirit to continued fractions. The fact that one can compute \texttt{reciprocal} trivially with continued fractions, may indicate that by changing the representation again to a continued fraction, we may improve the accuracy of $\tan{}$ and $\cot{}$ significantly.

It might be interesting to extend relational arithmetic of \cite{kiselyov2008pure}, which works using unary arithmetic, to binary and augment it with an implementation of \texttt{sqrt} as a relation. This can then be used to define a single relation implementing both forward-mode and reverse-mode trigonometric functions.

As of now, presented functions do not handle non-dyadic rationals or flonums. An obvious extension is to generalise presented method to a low-order Taylor expansion centered around the ``nearest'' dyadic rational. Since trigonometric functions are \emph{analytic} (viz., infinitely differentiatable), and because we can simultaneously compute both $\cos{}$ as well as its derivative ($-\sin{}$) by simple swapping of signs outside and inside the first radical, the Taylor theorem may be applied to yield a good approximation using only two or three terms.

I would like to thank the anonymous reviewers.

\clearpage

\begin{small}
\bibliographystyle{eptcs}
\bibliography{all.UTF-8}
\end{small}

\appendix

\section{Supplementary material}
\label{Supplementary material}

\begin{figure}[p]
\label{Helper macros}
\fbox{\hbox to .99\columnwidth{\vbox{\vspace{-.5em}{\small{\begin{flushleft}
{
{\let\oldtbf=\textbf\def\textbf#1{{\rm\oldtbf{#1}}}\noindent \texttt{\begin{tabular*}{0.90\linewidth}{l@{\extracolsep{\fill}}}
\textit{;; syntax-level λ abstraction}\\
(\textbf{def-syntax}\ {\char91}Λ\ forms\ .\ body{\char93}\\
\ \ \ (\textbf{syntax-rules}\ ..\ ()\ ({\char91}\_\ .\ forms{\char93}\ .\ body)))\\
\textit{;; untangle a concatenated list into 2 lists}\\
\textit{;; of elements (before and after the separator)}\\
\textit{;; (split (fn {\char91}{\char93} {\char91}{\char93} . rest) a b c ... SEP x y z ...)}\\
\textit{;; → (fn (a b c ...) (x y z ...) . rest)}\\
(\textbf{def-syntax}\ split\ (\textbf{syntax-rules}\ (SEP)\\
\ ({\char91}\_\ (k\ fst\ {\char91}snd\ ...{\char93}\ .\ r)\ SEP\ \ .\ rest{\char93}\\
\ \ (k\ fst\ {\char91}snd\ ...\ .\ rest{\char93}\ .\ r))\\
\ ({\char91}\_\ (k\ {\char91}fst\ ...{\char93}\ snd\ .\ r)\ x\ .\ rest{\char93}\\
\ \ (split\ (k\ {\char91}fst\ ...\ x{\char93}\ snd\ .\ r)\ .\ rest))\\
\ ))\\
\textit{;; Extract a single colored keyword. source: {\char91}Kis02{\char93}}\\
(\textbf{def-syntax}\ {\char91}extract\ sym\ body\ \_k{\char93}\\
\ \ \ (\textbf{letrec-syntax}\ ({\char91}tr\ (\textbf{syntax-rules}\ ..\ (sym)\\
\ \ \ \ \ ({\char91}\_\ x\ sym\ tail\ (k\ (syms\ ..)\ .\ args){\char93}\\
\ \ \ \ \ \ (k\ (syms\ ..\ x)\ .\ args))\\
\ \ \ \ \ ({\char91}\_\ d\ (x\ .\ y)\ tail\ k{\char93}\\
\ \ \ \ \ \ (tr\ x\ x\ (y\ .\ tail)\ k))\\
\ \ \ \ \ ({\char91}\_\ d1\ d2\ ()\ (k\ (syms\ ..)\ .\ args){\char93}\\
\ \ \ \ \ \ (k\ (syms\ ..\ sym)\ .\ args))\\
\ \ \ \ \ ({\char91}\_\ d1\ d2\ (x\ .\ y)\ k{\char93}\ (tr\ x\ x\ y\ k))){\char93})\\
\ \ \ \ \ \ (tr\ body\ body\ ()\ \_k)\\
\ \ \ ))\\
\textit{;; Extract multiple colored keywords. source: {\char91}Kis02{\char93}}\\
(\textbf{def-syntax}\ extract*\ (\textbf{syntax-rules}\ ()\\
\ \ \ \ ({\char91}\_\ (sym)\ body\ k{\char93}\ (extract\ sym\ body\ k))\\
\ \ \ \ ({\char91}\_\ \_syms\ \_body\ \_k{\char93}\\
\ \ \ \ \ (\textbf{letrec-syntax}\ ({\char91}ex-aux\\
\ \ \ \ \ \ \ (\textbf{syntax-rules}\ ()\\
\ \ \ \ \ \ \ \ \ \ ({\char91}\_\ syms\ ()\ body\ (k\ ()\ .\ args){\char93}\\
\ \ \ \ \ \ \ \ \ \ \ (k\ syms\ .\ args))\\
\ \ \ \ \ \ \ \ \ \ ({\char91}\_\ syms\ (sym\ .\ sym*)\ body\ k{\char93}\\
\ \ \ \ \ \ \ \ \ \ \ (extract\ sym\ body\\
\ \ \ \ \ \ \ \ \ \ \ \ \ \ (ex-aux\ syms\ sym*\ body\ k)))\\
\ \ \ \ \ \ \ \ \ \ ){\char93})\\
\ \ \ \ \ \ \ (ex-aux\ ()\ \_syms\ \_body\ \_k)))\\
))\end{tabular*}
}}
}\end{flushleft}
}}\vspace{-.5em}}}}
\caption{{\color{black}\label{helpers}
}{\color{black}}Helper macros}
\end{figure}

\subsection{Tangential: anatypes}
\label{Tangential: anatypes}

\begin{figure}[p]
\label{Defining anaphoric macros}
\fbox{\hbox to .99\columnwidth{\vbox{\vspace{-.5em}{\small{\begin{flushleft}
{
{\let\oldtbf=\textbf\def\textbf#1{{\rm\oldtbf{#1}}}\noindent \texttt{\begin{tabular*}{0.90\linewidth}{l@{\extracolsep{\fill}}}
(\textbf{def-syntax}\ {\char91}define-anaphora\ (thy\ ...)\ (n\ rd\ rl)\ ...{\char93}\\
\ \ (\textbf{let-syntax-rule}\ ({\char91}K\ (rb)\ ns\ ({\char91}name\ rdef\ rules{\char93}\ ..){\char93}\\
\ \ \ \ (\textbf{letrec-syntax}\ ({\char91}rdef\ rules{\char93}\ ..)\\
\ \ \ \ \ \ (\textbf{begin}\ (\textbf{def-syntax}\ {\char91}name\ .\ fs{\char93}\\
\ \ \ \ \ \ \ \ \ \ \ (extract*\ (thy\ ...\ SEP\ .\ ns)\ fs\\
\ \ \ \ \ \ \ \ \ \ \ \ \ {\char91}(Λ\ (al\ body)\\
\ \ \ \ \ \ \ \ \ \ \ \ \ \ \ (split\ (rdef\ {\char91}{\char93}\ {\char91}{\char93}\ .\ body)\ .\ al))\\
\ \ \ \ \ \ \ \ \ \ \ \ \ \ {\char91}{\char93}\ fs{\char93}\\
\ \ \ \ \ \ \ \ \ \ \ \ ))\ ..\\
\ \ \ \ \ \ \ \ )))\\
\ \ \ \ (extract*\ (recursive-bindings)\ (rl\ ...)\\
\ \ \ \ \ \ \ {\char91}K\ {\char91}{\char93}\ (n\ ...)\ ({\char91}n\ rd\ rl{\char93}\ ...){\char93})\\
\ \ ))\end{tabular*}
}}
}\end{flushleft}
}}\vspace{-.5em}}}}
\caption{{\color{black}\label{top}
}{\color{black}}Defining anaphoric macros}
\end{figure}

Often enough, one wants to specialise Scheme code to the types at hand. Although the language itself is uni-typed (with run-time type information attached to every variable), the hygienic \texttt{syntax-rules} macro system actually provides the means to create an annotation for every binding and secretly pass around syntactic information that can be used to statically overload common arithmetic operators and library functions. Other functions dealing with character, collections (i.e., vectors, structs, pairs, lists, strings, arrays and matrices) are not used in the paper, but still are supported. Our \texttt{sqrt-loop} could shed some clutter indeed if we could drop all type-specialization suffixes from the code.

\begin{figure}[p]
\label{Re-definition engine}
\fbox{\hbox to .99\columnwidth{\vbox{\vspace{-.5em}{\small{\begin{flushleft}
{
{\let\oldtbf=\textbf\def\textbf#1{{\rm\oldtbf{#1}}}\noindent \texttt{\begin{tabular*}{0.90\linewidth}{l@{\extracolsep{\fill}}}
(\textbf{def-syntax}\ redefine-thyself*\ (\textbf{syntax-rules}\ ()\\
\ ({\char91}\_\ {\color{red}"internal"}\ ids\ (slf\ ...)\ (rec\ ...)\ .\ body{\char93}\\
\ \ (\textbf{letrec-syntax}\ ({\char91}slf\ (Λ\ ts\\
\ \ \ (extract*\ ids\ ts\\
\ \ \ \ \ \ \ {\char91}(Λ\ (al\ terms)\\
\ \ \ \ \ \ \ \ \ \ \ (split\ (rec\ {\char91}{\char93}\ {\char91}{\char93}\ .\ terms)\ .\ al))\\
\ \ \ \ \ \ \ \ {\char91}{\char93}\ ts{\char93})\\
\ \ \ ){\char93}\ ...)\ .\ body\\
\ \ ))\\
\ ({\char91}\_\ (ids\ ...)\ {\char91}{\char93}\ (slf\ ...)\ recs\ .\ body{\char93}\\
\ \ (redefine-thyself*\ {\color{red}"internal"}\ (ids\ ...\ SEP\ slf\ ...)\\
\ \ \ \ \ (slf\ ...)\ recs\ .\ body))\\
\ ))\end{tabular*}
}}
}\end{flushleft}
}}\vspace{-.5em}}}}
\caption{{\color{black}\label{redef}
}{\color{black}}Re-definition engine}
\end{figure}

It seems natural to use a quoted data-like construct (viz., \emph{keywords} in LISP) for the annotation itself, and introduce a new set of basic binding constructs that support such annotations. These ``symbols'' are matched literally by the macro system, simplifying the expression of our macros somewhat. For the rest of the paper we shall primarily use new forms of \texttt{Define}, parallel \texttt{Let}, sequential \texttt{Let*}.
The supporting site \cite{Anaphora} includes implementations of many more standard and some less standard binding forms, all of which can be derived from the core forms of \texttt{Lambda} (introducing an annotated $\lambda$) and \texttt{As} (merely annotating an existing, possibly top-level binding).
In Figure \ref{cora} we give a definition of these two core forms, which makes use of \texttt{define-anaphora} machinery (implementation of which is somewhat involved but is described in Figures \ref{top} and \ref{redef}).

\begin{figure*}[p]\begin{multicols}{2}
\label{Core anatype forms (abridged)}
\end{multicols}\footnotesize\fbox{\hbox to .485\columnwidth{\vbox{\vspace{-.5em}{{\begin{flushleft}
{
{\let\oldtbf=\textbf\def\textbf#1{{\rm\oldtbf{#1}}}\noindent \texttt{\begin{tabular*}{0.35\columnwidth}{l@{\extracolsep{\fill}}}
(define-anaphora\ (type\ \textbf{+}\ \textbf{-}\ \textbf{*}\ \textbf{/}\ \textbf{=}\ <\ >\ <=\ >=\\
\ \ \ \textmd{\textit{neg}}\ \textmd{\textit{min}}\ \textmd{\textit{max}}\ \textmd{\textit{zero?}}\ \textmd{\textit{even?}}\ \textmd{\textit{odd?}}\ \textmd{\textit{positive?}}\\
\ \ \ \textmd{\textit{negative?}}\ \textbf{and}\ \textmd{\textit{lsh}}\ \textbf{not}\ \textbf{or}\ \textmd{\textit{rsh}}\ \textmd{\textit{xor}}\ \textmd{\textit{fix}}\ \textmd{\textit{flo}})\\
(\textbf{As}\ rec3\ (\textbf{syntax-rules}\ ()\\
\ ({\char91}\_\ {\color{red}"int"}\ os\ ss\ ct\ ()\ .\ k{\char93}\ (\textbf{begin}\ .\ k))\\
\ ({\char91}\_\ {\color{red}"int"}\ os\ ss\ ct\ (n\ m\ .\ r)\ .\ ts{\char93}\\
\ \ (TYPES\ (Λ\ types\ (member??\ n\ types\\
\ \ \ (rec3\ {\color{red}"int"}\ os\ ss\ n\ (m\ .\ r)\ .\ ts)\\
\ \ \ (member??\ m\ types\\
\ \ \ \ \ \ (rec3\ {\color{red}"int"}\ os\ ss\ ct\ r\\
\ \ \ \ \ \ \ \ (ty-sugar\ m\ n\ os\ \\
\ \ \ \ \ \ \ \ \ \ \ (redefine-thyself*\ os\ {\char91}{\char93}\ ss\\
\ \ \ \ \ \ \ \ \ \ \ \ \ \ recursive-bindings\ .\ ts)))\\
\ \ \ \ \ \ (rec3\ {\color{red}"int"}\ os\ ss\ ct\ (m\ .\ r)\\
\ \ \ \ \ \ \ \ (ty-sugar\ ct\ n\ os\ \\
\ \ \ \ \ \ \ \ \ \ \ (redefine-thyself*\ os\ {\char91}{\char93}\ ss\\
\ \ \ \ \ \ \ \ \ \ \ \ \ \ recursive-bindings\ .\ ts))))))))\\
\ ({\char91}\_\ {\color{red}"int"}\ os\ ss\ ct\ (n\ .\ r)\ .\ ts{\char93}\\
\ \ (TYPES\ (Λ\ types\ (member??\ n\ types\\
\ \ \ (rec3\ {\color{red}"int"}\ os\ ss\ n\ r\ .\ ts)\\
\ \ \ (rec3\ {\color{red}"int"}\ os\ ss\ ct\ r\\
\ \ \ \ \ \ (ty-sugar\ ct\ n\ os\ \\
\ \ \ \ \ \ \ \ (redefine-thyself*\ os\ {\char91}{\char93}\ ss\\
\ \ \ \ \ \ \ \ \ \ \ recursive-bindings\ .\ ts)))))))\\
\ ({\char91}\_\ {\color{red}"int"}\ \_\ .\ r{\char93}\ (syntax-error\ \textbf{As}\ .\ r))\ \ \ \ \ \ \ \ \\
\ ({\char91}\_\ os\ ss\ ns\ .\ body{\char93}\\
\ \ (rec3\ {\color{red}"int"}\ os\ ss\ :unknown\ ns\ .\ body))\\
))\\
\end{tabular*}
}}
}\end{flushleft}
}}\vspace{-.5em}}}}\fbox{\hbox to .485\columnwidth{\vbox{\vspace{-.5em}{{\begin{flushleft}
{
{\let\oldtbf=\textbf\def\textbf#1{{\rm\oldtbf{#1}}}\noindent \texttt{\begin{tabular*}{0.35\columnwidth}{l@{\extracolsep{\fill}}}
(\textbf{Lambda}\ rec4\ (\textbf{syntax-rules}\ ()\\
({\char91}\_\ {\color{red}"int"}\ os\ ss\ ct\ ()\ .\ k{\char93}\ k)\\
({\char91}\_\ {\color{red}"int"}\ os\ ss\ ct\ (n\ m\ .\ r)\ lt\ (b\ ...)\ .\ ts{\char93}\\
\ (TYPES\ (Λ\ types\ (member??\ n\ types\\
\ \ (rec4\ {\color{red}"int"}\ os\ ss\ n\ (m\ .\ r)\ lt\ (b\ ...)\ .\ ts)\\
\ \ (member??\ m\ types\\
\ \ \ \ \ (rec4\ {\color{red}"int"}\ os\ ss\ ct\ r\\
\ \ \ \ \ \ \ \ lt\ (b\ ...\ n)\\
\ \ \ \ \ \ \ \ (ty-sugar\ m\ n\ os\ \\
\ \ \ \ \ \ \ \ \ \ \ (redefine-thyself*\ os\ {\char91}{\char93}\ ss\\
\ \ \ \ \ \ \ \ \ \ \ \ \ \ recursive-bindings\ .\ ts)))\\
\ \ \ \ \ (rec4\ {\color{red}"int"}\ os\ ss\ ct\ (m\ .\ r)\\
\ \ \ \ \ \ \ \ lt\ (b\ ...\ n)\\
\ \ \ \ \ \ \ \ (ty-sugar\ ct\ n\ os\ \\
\ \ \ \ \ \ \ \ \ \ \ (redefine-thyself*\ os\ {\char91}{\char93}\ ss\\
\ \ \ \ \ \ \ \ \ \ \ \ \ \ recursive-bindings\ .\ ts))))))))\\
({\char91}\_\ {\color{red}"int"}\ os\ ss\ ct\ (n\ .\ r)\ lt\ (b\ ...)\ .\ ts{\char93}\\
\ (TYPES\ (Λ\ types\ (member??\ n\ types\\
\ \ (rec4\ {\color{red}"int"}\ os\ ss\ n\ r\ lt\ (b\ ...)\ .\ ts)\\
\ \ (rec4\ {\color{red}"int"}\ os\ ss\ ct\ r\\
\ \ \ \ \ lt\ (b\ ...\ n)\\
\ \ \ \ \ (ty-sugar\ ct\ n\ os\ \\
\ \ \ \ \ \ \ \ (redefine-thyself*\ os\ {\char91}{\char93}\ ss\\
\ \ \ \ \ \ \ \ \ \ \ recursive-bindings\ .\ ts)))))))\\
({\char91}\_\ {\color{red}"int"}\ \_\ .\ r{\char93}\ (syntax-error\ \textbf{Lambda}\ .\ r))\ \ \ \ \ \ \ \ \ \ \ \ \ \\
({\char91}\_\ os\ ss\ nvs\ .\ body{\char93}\\
\ (rec4\ {\color{red}"int"}\ os\ ss\ :unknown\ nvs\ λ\ ()\ .\ body))\\
))\ \textit{;; ... other anatype binding forms ...) }\\
\end{tabular*}
}}
}\end{flushleft}
}}\vspace{-.5em}}}}
\caption{{\color{black}\label{cora}
}{\color{black}}Core anatype forms (abridged)}
\end{figure*}

The principle of operation can be described as ``infectious'' spreading of syntactic information across binding scopes (with credits to Oleg Kiselyov \cite{Kiselyov02a} for the term). The same symbol may be annotated differently in different scopes. Each occurrence of this symbol will therefore cause overloaded operators or library functions to (syntactically) expand to their anatype-specialised versions.

\begin{figure}[p]
\label{Anatype support macros}
\fbox{\hbox to .99\columnwidth{\vbox{\vspace{-.5em}{\small{\begin{flushleft}
{
{\let\oldtbf=\textbf\def\textbf#1{{\rm\oldtbf{#1}}}\noindent \texttt{\begin{tabular*}{0.90\linewidth}{l@{\extracolsep{\fill}}}
\textit{;; all supported anatypes}\\
(\textbf{def-syntax}\ {\char91}TYPES\ k{\char93}\ (k\ :unknown\ \textmd{\textit{:fix}}\ \textmd{\textit{:flo}}))\\
\textit{;; base-case for type}\\
(\textbf{def-syntax}\ {\char91}type\ n\ k{\char93}\ (k\ :unknown))\\
\textit{;; deliver the anatype to the continuation}\\
(\textbf{def-syntax}\ {\char91}ty-trf\ type\ typ\ n{\char93}\ (\textbf{syntax-rules}\ (n)\\
\ \ \ \ \ \ ({\char91}\_\ n\ k{\char93}\ (k\ type))\\
\ \ \ \ \ \ ({\char91}\_\ .\ r{\char93}\ (typ\ .\ r))\\
))\\
\textit{;; overload an operator for x}\\
(\textbf{def-syntax}\ {\char91}op-trf\ opo\ opd\ x{\char93}\\
\ \ (\textbf{letrec-syntax}\ ({\char91}op\ (\textbf{syntax-rules}\ ..\ (x)\\
\ \ \ \ \ ({\char91}\_\ {\color{red}"internal"}\ (c\ ..)\ x\ .\ y{\char93}\ (opo\ c\ ..\ x\ .\ y))\\
\ \ \ \ \ ({\char91}\_\ {\color{red}"internal"}\ (c\ ..)\ y\ .\ z{\char93}\ (op\ {\color{red}"internal"}\ (c\ ..\ y)\ .\ z))\\
\ \ \ \ \ ({\char91}\_\ {\color{red}"internal"}\ c{\char93}\ (opd\ .\ c))\\
\ \ \ \ \ ({\char91}\_\ .\ rest{\char93}\ (op\ {\color{red}"internal"}\ {\char91}{\char93}\ .\ rest))){\char93})\\
\ \ \ (\textbf{syntax-rules}\ (op)\\
\ \ \ \ \ ({\char91}\_\ .\ args{\char93}\ (op\ .\ args)))\\
\ \ \ ))\\
\textit{;; base-case for ty-sugar}\\
(\textbf{def-syntax}\ {\char91}ty-sugar\ .\ args{\char93}\ (syntax-error\ .\ args))\\
\textit{;; redefine ty-sugar in a modular fashion}\\
(\textbf{def-syntax}\ {\char91}extend-ty-sugar\ clause\ ...{\char93}\\
\ \ (\textbf{def-syntax}\ ty-sugar\ (\textbf{let-syntax}\ ({\char91}tysugar\ ty-sugar{\char93})\\
\ \ \ \ (\textbf{syntax-rules}\ ()\ clause\ ...\\
\ \ \ \ \ \ \ ({\char91}\_\ .\ args{\char93}\ (tysugar\ .\ args))\\
\ \ \ \ \ \ \ ))\\
\ \ \ \ \ ))\\
\textit{;; Comparing symbols for syntactic equality}\\
(\textbf{def-syntax}\ {\char91}eq??\ th\ t\ kt\ kf{\char93}\\
\ \ \ ((\textbf{syntax-rules}\ (th)\\
\ \ \ \ \ \ \ \ ({\char91}\_\ th{\char93}\ kt)\\
\ \ \ \ \ \ \ \ ({\char91}\_\ tt{\char93}\ kf)\\
\ \ \ \ \ \ \ \ )\ t))\\
\textit{;; get the n'th form from the db according to spec}\\
(\textbf{def-syntax}\ {\char91}nth\ db\ .\ spec{\char93}\\
\ \ (\textbf{letrec-syntax}\ ({\char91}rec\ (\textbf{syntax-rules}\ ..\ ()\\
\ \ \ \ \ \ \ \ \ \ \ \ \ \ \ \ \ \ \ \ \ \ \ \ \ \ ({\char91}\_\ (y\ ..\ x){\char93}\ x)\\
\ \ \ \ \ \ \ \ \ \ \ \ \ \ \ \ \ \ \ \ \ \ \ \ \ \ ({\char91}\_\ (y\ ..\ x)\ 1\ .\ r{\char93}\ (rec\ (y\ ..)\ .\ r))\\
\ \ \ \ \ \ \ \ \ \ \ \ \ \ \ \ \ \ \ \ \ \ \ \ \ \ ){\char93})\\
\ \ \ \ \ (db\ (rec\ {\char91}'{\char35}\textbf{?}{\char93}\ .\ spec))\\
\ \ \ \ \ ))\\
\end{tabular*}
}}
}\end{flushleft}
}}\vspace{-.5em}}}}
\caption{{\color{black}\label{supp}
}{\color{black}}Anatype support macros}
\end{figure}

The core of the anatype implementation for overloading therefore contains a modular, extensible engine for resolving operators and library function bindings to their ana-typed variants. Starting from a stub for an \texttt{unknown:} anatype, each module may add new ``sugar'' to the syntactic environment, automatically causing all symbols annotated as e.g., \texttt{fix:} or \texttt{flo:} to be treated as fixnums or flonums, respectively. Figure \ref{arith} specifies some mappings that can be specialised this way, using separate syntax transformers \texttt{ty-trf} and \texttt{op-trf} (see Figure \ref{supp} also for the implementation of \texttt{extend-ty-sugar}).

\begin{figure*}[p]\begin{multicols}{2}
\label{Arithmetic anatypes (abridged)}
\end{multicols}\footnotesize\fbox{\hbox to .485\columnwidth{\vbox{\vspace{-.5em}{{\begin{flushleft}
{
{\let\oldtbf=\textbf\def\textbf#1{{\rm\oldtbf{#1}}}\noindent \texttt{\begin{tabular*}{0.90\linewidth}{l@{\extracolsep{\fill}}}
(extend-ty-sugar\ (\textmd{\textit{:fix}})\ \ \textit{;; fixnums}\\
({\char91}\_\ \textmd{\textit{:fix}}\ n\ (ty\ o+\ o-\ o*\ o/\ \textbf{=}\ <\ >\ <=\ >=\ \textmd{\textit{neg}}\\
\ \ \textmd{\textit{min}}\ \textmd{\textit{max}}\ \textmd{\textit{zero?}}\ \textmd{\textit{even?}}\ \textmd{\textit{odd?}}\ \textmd{\textit{positive?}}\\
\ \ \textmd{\textit{negative?}}\ \textbf{and}\ \textmd{\textit{lsh}}\ \textbf{not}\ \textbf{or}\ \textmd{\textit{rsh}}\ \textmd{\textit{xor}}\ \textmd{\textit{length}}\ ...\\
\ \ tofix\ toflo)\ .\ rs{\char93}\\
(\textbf{let-syntax}\ ({\char91}ty\ (ty-trf\ \textmd{\textit{:fix}}\ ty\ n){\char93}\\
\ {\char91}o+\ (op-trf\ +fx\ o+\ n){\char93}\ {\char91}o-\ (op-trf\ -fx\ o-\ n){\char93}\\
\ ...\ other\ overloads\ ...)\ .\ rs)\\
))\end{tabular*}
}}
}\end{flushleft}
}}\vspace{-.5em}}}}\fbox{\hbox to .485\columnwidth{\vbox{\vspace{-.5em}{{\begin{flushleft}
{
{\let\oldtbf=\textbf\def\textbf#1{{\rm\oldtbf{#1}}}\noindent \texttt{\begin{tabular*}{0.90\linewidth}{l@{\extracolsep{\fill}}}
(extend-ty-sugar\ (\textmd{\textit{:flo}})\ \ \textit{;; flonums}\\
({\char91}\_\ \textmd{\textit{:flo}}\ n\ (ty\ o+\ o-\ o*\ o/\ \textbf{=}\ <\ >\ <=\ >=\ \textmd{\textit{neg}}\\
\ \ \textmd{\textit{min}}\ \textmd{\textit{max}}\ \textmd{\textit{zero?}}\ \textmd{\textit{even?}}\ \textmd{\textit{odd?}}\ \textmd{\textit{positive?}}\\
\ \ \textmd{\textit{negative?}}\ tofix\ toflo)\ .\ rs{\char93}\\
(\textbf{let-syntax}\ ({\char91}ty\ (ty-trf\ \textmd{\textit{:flo}}\ ty\ n){\char93}\\
\ {\char91}o+\ (op-trf\ +fl\ o+\ n){\char93}\ {\char91}o-\ (op-trf\ -fl\ o-\ n){\char93}\\
\ ...\ other\ overloads\ ...)\ \\
.\ rs)\\
))\end{tabular*}
}}
}\end{flushleft}
}}\vspace{-.5em}}}}
\caption{{\color{black}\label{arith}
}{\color{black}}Arithmetic anatypes (abridged)}
\end{figure*}

One may wonder how this is implemented in pure Scheme, as the definition of \texttt{+}, for example, seems to be subjected to non-lexical scoping rules. The answer is simple: weak hygiene. Already described by Oleg Kiselyov, the idea is convert macros to \mac{CPS} \cite{Kiselyov02} and then subvert weak hygiene for locally introduced bindings using Petrofsky extraction. Subsequently the binding macros themselves are re-defined such that any nested re-definition of the binding that is subjected to such ``hygiene breaking'' is properly propagated (i.e., ``infected'') into deeper scopes. More background details of the technique can be found in \cite{Kiselyov02a}.

All this allows us to express the \texttt{sqrt-loop/sqrt-start} and all our trigonometric functions in much more succinct way. These are, of course not a replacement for regular types: there is neither built-in type-checking nor inference. Anaphoric types serve as mere annotations, supporting overloading in a seamless and transparent way.

{\small{\begin{flushleft}
{
{\let\oldtbf=\textbf\def\textbf#1{{\rm\oldtbf{#1}}}\noindent \texttt{\begin{tabular*}{0.90\linewidth}{l@{\extracolsep{\fill}}}
\\
(\textbf{Define}\ (sqrt-loop\ fix:\ n\ start)\\
\ \ \ (\textbf{Let}\ loop\ (fix:{\char91}num\ n{\char93}\\
\ \ \ \ \ \ \ \ \ \ \ \ \ \ \ \ \ \ {\char91}res\ 0{\char93}\\
\ \ \ \ \ \ \ \ \ \ \ \ \ \ \ \ \ \ {\char91}bit\ start{\char93})\\
\ \ \ \ \ \ (\textbf{cond}\ ({\char91}\textmd{\textit{zero?}}\ bit{\char93}\ res)\\
\ \ \ \ \ \ \ \ \ \ \ \ ({\char91}>=\ num\ (\textbf{+}\ res\ bit){\char93}\\
\ \ \ \ \ \ \ \ \ \ \ \ \ (loop\ (\textbf{-}\ num\ (\textbf{+}\ res\ bit))\\
\ \ \ \ \ \ \ \ \ \ \ \ \ \ \ \ (\textbf{+}\ bit\ (\textmd{\textit{rsh}}\ res\ 1))\\
\ \ \ \ \ \ \ \ \ \ \ \ \ \ \ \ (\textmd{\textit{rsh}}\ bit\ 2)))\\
\ \ \ \ \ \ \ \ \ \ \ \ (\textbf{else}\\
\ \ \ \ \ \ \ \ \ \ \ \ \ (loop\ num\ (\textmd{\textit{rsh}}\ res\ 1)\ (\textmd{\textit{rsh}}\ bit\ 2)))\\
\ \ \ )))\\
\\
\end{tabular*}
}}
}\end{flushleft}
}}

\subsection{Tangential: threading}
\label{Tangential: threading}

We could stop here - the algorithm is clear enough. The abundance of parentheses still hurts the eye, however, so let us deploy our next secret weapon. Threading introduces an explicit \emph{anaphoric} keyword (\texttt{\_} below) that binds successive terms that are arranged in a sequence (i.e., a thread) in 3 possible ways: syntactically (with \texttt{>->}), using a $\lambda$ (with \texttt{>=>}) aka \mac{ANF} or using a \textsc{Maybe} monad (using Scheme's \texttt{{\char35}f} as \texttt{Nothing}). The latter two are not used in this paper, and thus are elided from Figure \ref{threading}. These definitions are trivial to derive from the given definition of \texttt{>->}.

\begin{figure}[p]
\label{define-anaphora for (syntactic) threading}
\fbox{\hbox to .99\columnwidth{\vbox{\vspace{-.5em}{\small{\begin{flushleft}
{
{\let\oldtbf=\textbf\def\textbf#1{{\rm\oldtbf{#1}}}\noindent \texttt{\begin{tabular*}{0.90\linewidth}{l@{\extracolsep{\fill}}}
(\textbf{def-syntax}\ decimal->unary\ (\textbf{syntax-rules}\ ()\\
\ ({\char91}\_\ k\ 0{\char93}\ (k))\ \ \ \ \ \ \ \ \ \ \ \ \textit{;; syntactically convert decimal}\\
\ ({\char91}\_\ k\ 1{\char93}\ (k\ 1))\ \ \ \ \ \ \ \ \ \ \textit{;; to unary for a reasonable}\\
\ ({\char91}\_\ k\ 2{\char93}\ (k\ 1\ 1))\ ...\ ))\ \textit{;; number of n, e.g., 0..10}\\
(define-anaphora\ (\_\ \_\_)\\
\ \ (>->\ rec1\ (\textbf{syntax-rules}\ ()\\
\ \ \ ({\char91}\_\ \_\ x\ fa{\char93}\ fa)\\
\ \ \ ({\char91}\_\ (thy\ pre)\ slfs\ fa\ .\ fb{\char93}\\
\ \ \ \ (\textbf{let-syntax}\ ({\char91}thy\ fa{\char93})\\
\ \ \ \ \ \ (\textbf{let-syntax-rule}\ ({\char91}pre\ (k\ db\ .\ as){\char93}\\
\ \ \ \ \ \ \ \ \ \ \ \ \ \ \ \ \ \ \ \ \ (pre\ (k\ (thy\ .\ db)\ .\ as)))\\
\ \ \ \ \ \ \ \ (redefine-thyself*\ (thy\ pre)\ {\char91}{\char93}\ slfs\ (rec1\ nrec)\\
\ \ \ \ \ \ \ \ \ \ \ (rec1\ (thy\ pre)\ slfs\ .\ fb)\\
\ \ \ \ \ \ \ \ \ \ \ ))))))\\
\ \ (\textbf{Nth}\ nrec\ (\textbf{syntax-rules}\ ()\\
\ \ \ ({\char91}\_\ (thy\ pre)\ slfs\ n{\char93}\\
\ \ \ \ (decimal->unary\ (Λ\ nl\ (nth\ pre\ .\ nl))\ n)))\\
\ \ ))\\
\end{tabular*}
}}
}\end{flushleft}
}}\vspace{-.5em}}}}
\caption{{\color{black}\label{threading}
}{\color{black}}define-anaphora for (syntactic) threading}
\end{figure}

Similar to anaphoric types, threading also touches upon weak hygiene: the anaphoric keyword is bound differently and implicitly for each sub-expression. For example, calculation of the next value of \texttt{sig} can be described in English in the following sentence. First, take the value of \texttt{sig}, shift it to the left, then ``or'' it with a new bit depending on the bit-count of \texttt{sig} (refer to the beginning of the sentence which is 2 levels before this one). Alternatively, the same sentence can be expressed using threading:

{\small{\begin{flushleft}
{
{\let\oldtbf=\textbf\def\textbf#1{{\rm\oldtbf{#1}}}\noindent \texttt{\begin{tabular*}{0.90\linewidth}{l@{\extracolsep{\fill}}}
(bit-or\ (\textmd{\textit{lsh}}\ s\ 1)\ \ \ \ \ \ \ \ \ \ \ \ \ \ \ \ \ \ \ \ \ \ \ \ \ \ \ \ \ \ \textit{;; s has fix: anatype}\\
\ \ \ \ \ \ \ \ (\textbf{if}\ (\textmd{\textit{even?}}\ (\textmd{\textit{bit-count}}\ s))\\
\ \ \ \ \ \ \ \ \ \ \ \ 1\ 0))\\
\end{tabular*}
}}
}\end{flushleft}
}}
\noindent $-\{$introduce threading$\}\rightarrow$
{\small{\begin{flushleft}
{
{\let\oldtbf=\textbf\def\textbf#1{{\rm\oldtbf{#1}}}\noindent \texttt{\begin{tabular*}{0.90\linewidth}{l@{\extracolsep{\fill}}}
(>->\ s\ \ \ \ \ \ \ \ \ \ \ \ \ \ \ \ \ \ \ \ \ \ \ \ \ \ \ \ \ \ \ \ \ \ \ \ \ \ \ \ \ \textit{;; s is already annotated with fix:}\\
\ \ \ \ \ (\textmd{\textit{lsh}}\ \_\ 1)\ \ \ \ \ \ \ \ \ \ \ \ \ \ \ \ \ \ \ \ \ \ \ \ \ \ \ \ \ \ \ \ \ \textit{;; sub-expressions inherit this anatype}\\
\ \ \ \ \ (\textbf{or}\ \_\ (\textbf{if}\ (\textmd{\textit{even?}}\ (\textmd{\textit{bit-count}}\ {\char91}\textbf{Nth}\ 2{\char93}))\\
\ \ \ \ \ \ \ \ \ \ \ \ \ \ \ 1\ 0)))\\
\end{tabular*}
}}
}\end{flushleft}
}}

The ability to refer to the previous sub-expression (or indeed to any n'th sub-expression preceding the current one in closest thread introduced by a \texttt{>->}) is realised using anaphoric macros too. The binding for \texttt{\_} needs to be shadowed in each scope introduced by a new sub-expression, and therefore the macro-redefinition machinery (see Figure \ref{redef}) introduced above for anatypes can be leveraged here as well. Figure \ref{threading} lists a representative implementation of this anaphoric construct.

\subsection{Tangential: rational arithmetic}
\label{Tangential: rational arithmetic}

Although the code seems to be working well, with no explicit squaring/multiplication operations involved, it still can be improved. One shortcoming is that dyadic rationals are not represented as such explicitly. Including exact rationals would improve presentation of the algorithm as well as support the use of exact arithmetic.

We can not present a complete module implementing rationals here (it would be boring anyway), so let us for example, take $\sqrt{n\over m}$. This function can be computed using our implementation from section \ref{The (square) root of all good} and is quite representative for our purposes (we assume \texttt{/} and other basic arithmetic operators have already been overloaded for rationals).

{\small{\begin{flushleft}
{
{\let\oldtbf=\textbf\def\textbf#1{{\rm\oldtbf{#1}}}\noindent \texttt{\begin{tabular*}{0.90\linewidth}{l@{\extracolsep{\fill}}}
(\textbf{def-syntax}\ *maxval*\ 1152921504606846975)\ \ \ \ \ \ \ \ \ \ \ \ \textit{;; (- (exp2 *maxbit*) 1)}\\
(\textbf{def-syntax}\ *maxstart*\ 288230376151711744)\ \ \ \ \ \ \ \ \ \ \ \textit{;; (exp2 (- *maxbit 2))}\\
(\textbf{Define}\ sqrt/rat\\
\ \ (\textbf{fn-with}\ (\textbf{apply}\ \textbf{sqrt})\ \ \ \ \ \ \ \ \ \ \ \ \ \ \ \ \ \ \ \ \ \ \ \ \ \ \ \ \ \ \textit{;; chain the number tower}\\
\ \ \ ❘\ (〈rat〉\ :\ n\ m)\ =>\\
\ \ \ \ (\textbf{Let*}\ (fix:{\char91}x\ (\textmd{\textit{max}}\ n\ m){\char93}\ \ \ \ \ \ \ \ \ \ \ \ \ \ \ \ \ \ \ \ \ \ \ \ \ \textit{;; determine the best}\\
\ \ \ \ \ \ \ \ \ \ \ \ \ \ \ {\char91}f\ (\textbf{/}\ *maxval*\ x){\char93}\ \ \ \ \ \ \ \ \ \ \ \ \ \ \ \ \ \ \ \ \textit{;; available precision}\\
\ \ \ \ \ \ \ \ \ \ \ \ \ \ \ {\char91}n\ (\textbf{*}\ n\ f){\char93}\ \ \ \ \ \ \ \ \ \ \ \ \ \ \ \ \ \ \ \ \ \ \ \ \ \ \ \textit{;; rescale n/m for}\\
\ \ \ \ \ \ \ \ \ \ \ \ \ \ \ {\char91}m\ (\textbf{*}\ m\ f){\char93})\ \ \ \ \ \ \ \ \ \ \ \ \ \ \ \ \ \ \ \ \ \ \ \ \ \ \textit{;; that precision }\\
\ \ \ \ (\textbf{/}\ (sqrt-loop\ n\ (sqrt-start\ n))\\
\ \ \ \ \ \ \ (sqrt-loop\ m\ (sqrt-start\ m))\\
\ \ \ \ \ \ \ ))\\
\ \ \ \ ))\\
(\textbf{def-syntax}\ \textbf{sqrt}\ sqrt/rat)\ \ \ \ \ \ \ \ \ \ \ \ \ \ \ \ \ \ \ \ \ \ \ \ \ \ \ \ \textit{;; overload sqrt for rationals}\\
\end{tabular*}
}}
}\end{flushleft}
}}

Since rational values that we denote in this paper using \texttt{{\char35}r:n/m} reader syntax are represented by Scheme records (with numerator \texttt{n} and denominator \texttt{m}) they can be matched using an extensible pattern-matcher \cite{Kourzanov15}. In case of a match failure, a default implementation of \texttt{sqrt} (specified as the first form of \texttt{fn-with} pattern-matching $\lambda$) is applied on the argument. Otherwise, both parts are re-scaled and passed to \texttt{sqrt-loop}. This ensures that the resulting rational represents the ``best'' rational approximation to the square root of the argument rational given maximal precision of available fixnums.

An interesting observation is that trigonometric functions such as \texttt{dyadic-cos1} only use basic operators that are exact for rationals (besides the above \texttt{sqrt}, of course). Therefore, passing an initial factor \texttt{f=(<rat> : 1 4)} gives the best rational approximation of the result automatically, with no change to the algorithm.

\subsection{Tangential: composition transformers}
\label{Tangential: composition transformers}

A very useful functional-programming technique that is not streamlined in Scheme as well as it is in e.g., Haskell is currying (aka sectioning) and composition of functions. Traditionally, both currying and composition in Scheme is performed at the value-level, leading to extraneous closures even when the argument was known. We chose to perform composition at the syntax-level, possibly translating to value-level when needed. For the purposes of this paper, the following macro-based higher-order transformations suffice:

{\small{\begin{flushleft}
{
{\let\oldtbf=\textbf\def\textbf#1{{\rm\oldtbf{#1}}}\noindent \texttt{\begin{tabular*}{0.90\linewidth}{l@{\extracolsep{\fill}}}
(\textbf{def-syntax}\ {\char91}∘\ .\ fs{\char93}\ \ \ \ \ \ \ \ \ \ \ \ \ \ \ \ \ \ \ \ \ \ \textit{;; => Value-level}\\
\ \ \ (Λ\ args\ \ \ \ \ \ \ \ \ \ \ \ \ \ \ \ \ \ \ \ \ \ \ \ \ \ \ \ \ \ \ \ \textit{;; => λ args}\\
\ \ \ \ \ \ (\textbf{letrec-syntax}\ ({\char91}compose\\
\ \ \ \ \ \ \ \ \ (\textbf{syntax-rules}\ ()\\
\ \ \ \ \ \ \ \ \ \ \ \ ({\char91}\_\ g{\char93}\ (g\ .\ args))\ \ \ \ \ \ \ \ \ \ \ \ \textit{;; => (apply g args)}\\
\ \ \ \ \ \ \ \ \ \ \ \ ({\char91}\_\ f\ .\ gs{\char93}\\
\ \ \ \ \ \ \ \ \ \ \ \ \ (f\ (compose\ .\ gs)))\\
\ \ \ \ \ \ \ \ \ \ \ \ ){\char93})\\
\ \ \ \ \ \ \ \ \ (compose\ .\ fs)\\
\ \ \ \ \ \ \ \ \ )))\\
\\
(\textbf{def-syntax}\ {\char91}κ\ .\ exps{\char93}\ \ \ \ \ \ \ \\
\ \ \ (extract*\ (\_)\ exps\ {\char91}\ \ \ \ \ \ \ \ \ \ \ \ \ \ \ \ \ \ \ \textit{;; extract keyword \_ in any scope}\\
\ \ \ \ \ \ \ \ \ (Λ\ ((x)\ terms)\ \ \ \ \ \ \ \ \ \ \ \ \ \ \ \ \ \ \ \textit{;; obtain colored \_ as x and terms}\\
\ \ \ \ \ \ \ \ \ \ \ (λ\ x\ \ terms))\ \ \ \ \ \ \ \ \ \ \ \ \ \ \ \ \ \ \textit{;; introduce a λ }\\
\ \ \ \ \ \ \ \ \ \ \ \ \ {\char91}{\char93}\ \ exps{\char93}\ \ \ \ \ \ \ \ \ \ \ \ \ \ \ \ \ \ \ \ \textit{;; initial values}\\
))\\
\end{tabular*}
}}
}\end{flushleft}
}}

So for example, point-free definitions of common operators can now be given very concisely:

{\small{\begin{flushleft}
{
{\let\oldtbf=\textbf\def\textbf#1{{\rm\oldtbf{#1}}}\noindent \texttt{\begin{tabular*}{0.90\linewidth}{l@{\extracolsep{\fill}}}
(\textbf{Define}\ \textbf{add1}\ {\char91}κ\ \textbf{+}\ 1\ \_\ {\char93})\ (\textbf{Define}\ \textbf{sub1}\ {\char91}κ\ \textbf{-}\ \_\ 1{\char93})\\
(\textbf{Define}\ \textbf{double}\ {\char91}κ\ \textbf{*}\ \_\ 2{\char93})\ (\textbf{Define}\ \textbf{half}\ {\char91}κ\ \textbf{/}\ \_\ 2{\char93})\\
(\textbf{Define}\ \textbf{reciprocal}\ {\char91}κ\ \textbf{/}\ 1\ \_\ {\char93})\ (\textbf{Define}\ \textbf{sqr}\ {\char91}κ\ \textbf{*}\ \_\ \_{\char93})\\
\end{tabular*}
}}
}\end{flushleft}
}}

\section{Acronyms}\label{sec:acros}
\begin{acronym}
\macro{ANF} {A-Normal Form}
\macro{CORDIC} {Coordinate Rotation Digital Computer}
\macro{CPS} {Continuation Passing Style}
\macro{DSL} {Domain-Specific Language}
\macro{FFT} {Fast Fourier Transform}
\macrod{GCC} {GNU Compiler Collection} {formerly GNU C Compiler}
\macro{ML} {Meta Language}
\macro{MXM} {Max Magnitude}
\macro{R5RS} {Revised$^5$ Report on the Algorithmic Language Scheme}
\macro{RMS} {Root Mean Square}
\macro{SRFI} {Scheme Request for Implementation}
\end{acronym}

\end{document}